\documentclass[9pt,a4paper]{extarticle}
\usepackage{float}
\usepackage[a4paper]{geometry}
\usepackage[utf8]{inputenc}
\usepackage{csquotes}
\usepackage{hyperref}
\hypersetup{pdfborder={0 0 0} } % This turns off the border around links
\usepackage{authblk}
\usepackage[
    backend=biber, 
    style=ieee, 
    citestyle=numeric-comp, 
    sortcites=true, 
    maxnames= 3, 
    minnames=2]{biblatex}
\usepackage{color}
\usepackage{graphicx}
\usepackage[font=footnotesize,labelfont=bf]{caption}
\usepackage[font=footnotesize,labelfont=bf]{subcaption}
\usepackage[T1]{fontenc}
\usepackage{palatino}
\usepackage{float}

\title{You Don't Have to Live Next to Me: Towards Demobilizing Individualistic Bias in Computational Approaches to Urban Segregation}

\date{\small July 2025}

\author[*1,2]{\large Anastassia Vybornova}
\author[3]{\large Trivik Verma}

\affil[1]{\small University of Copenhagen, SODAS (Copenhagen Center for Social Data Science)}
\affil[2]{\small IT University of Copenhagen, NERDS (NEtwoRks, Data, and Society)}
\affil[3]{\small University of Bristol}
\affil[*]{\small \href{mailto:anvy@itu.dk}{anvy@itu.dk}}

\begin{document}

\maketitle

\begin{abstract} 
\begin{small}
The global surge in social inequalities is one of the most pressing issues of our times. The spatial expression of social inequalities at city scale gives rise to urban segregation, a common phenomenon across different local and cultural contexts. The increasing popularity of Big Data and computational models has inspired a growing number of computational social science studies that analyze, evaluate, and issue policy recommendations for urban segregation. Today's wealth in information and computational power could inform urban planning for equity. However, as we show here, segregation research is epistemologically interdependent with prevalent economic theories which overfocus on individual responsibility while neglecting systemic processes. This individualistic bias is also engrained in computational models of urban segregation. Through several contemporary examples of how Big Data -- and the assumptions underlying its usage -- influence (de)segregation patterns and policies, our essay tells a cautionary tale. We highlight how a lack of consideration for data ethics can lead to the creation of computational models that have a real-life, further marginalizing impact on disadvantaged groups. With this essay, our aim is to develop a better discernment of the pitfalls and potentials of computational approaches to urban segregation, thereby fostering a conscious focus on systemic thinking about urban inequalities. We suggest setting an agenda for research and collective action that is directed at demobilizing individualistic bias, informing our thinking about urban segregation, but also more broadly our efforts to create sustainable cities and communities.
\end{small}
\end{abstract}

\clearpage

\tableofcontents

\clearpage

\section{Mathematics in the Ghetto}
\label{sec:mathematics}
When you hear the word ``segregation'', do you conjure a mental image of a run-down, deprived neighbourhood? Maybe one that you have lived in, passed through, or seen from above in aerial pictures \cite{combs_aerial_2019}? Perhaps, you have heard the unsettling and callous use of the term ``ghetto'' \cite{schwartz_how_2019} in association with present-day urban segregation. If you have lived in Amsterdam, you might associate segregation with street crime \cite{pinkster_stickiness_2020}. Growing up in Delhi might evoke memories of neighbourhoods segregated by religion \cite{jamil_accumulation_2017}. If you identify as Czech, you might think of the spatial exclusion of the Roma community in your natal city \cite{cernusakova_stigma_2020}. In the UK, the othering might be happening along citizenship status \cite{bartram_uk_2019, burnett_towards_2020}, while in the US, division is most often interpreted along racial lines \cite{bonilla-silva_racism_2006, pattillo_black_2010}. Irrespective of where home might be, examples of local context run the gamut. Unless you happen to be an old-school casteist, racist, or classist, you might even bemoan the fact that your city is segregated, and wonder what could be done about it. Precisely because urban segregation is so common across different cultural contexts, it is a subject we can all relate to based on personal experiences. As often happens with subjects related to politics of place-making, it is also an emotionally charged one, and merits handling with utmost care.

\subsubsection*{The Schelling Model of Segregation}

\begin{figure}[ht]
\centering
\includegraphics[width=0.89\textwidth]{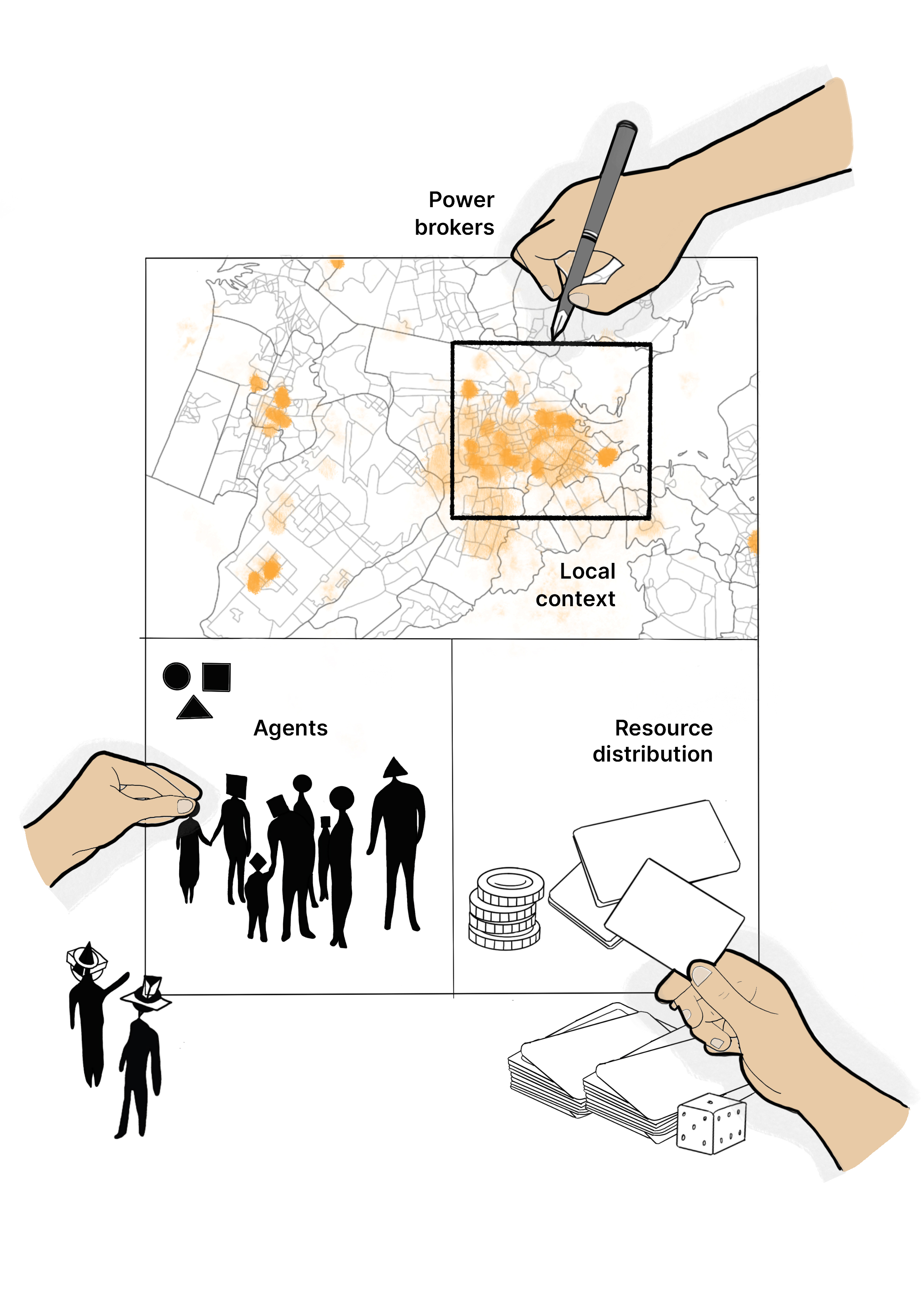}
\caption{The typical explicit and implicit building blocks of computational approaches to segregation: the spatial extent of the \textbf{local context} is outlined by \textbf{power brokers}. People with multifaceted identities (3D hats) are modelled as anonymous \textbf{agents} and reduced to single identity lines (2D hats), such as race or ethnicity. \textbf{Resource distribution} is skewed by definition, and gatekept by \textbf{power brokers}.}
\label{fig:mathematics}
\end{figure}

As academics focused on computation, our first introduction to systemic thinking about segregation was the so-called Schelling model \cite{schelling_models_1969, schelling_dynamic_1971}, one of the most influential agent-based models of segregation \cite{wolf_neighbourhood_2024} developed by Thomas C.~Schelling. The model suggests that even when individuals exhibit only mild preferences to live among others with similar traits, high levels of segregation can materialise. By engaging with the Schelling model, our experiences of facing segregation within our neighbourhoods and schools were embedded in a broader context, and the corresponding computational tools (see Figure~\ref{fig:mathematics}) enabled us to attach computationally simple and enticingly counter-intuitive explanations to our lived realities. 

Then, a few years ago, one of the authors of this work attended a Complex Systems lecture on the Schelling model at a university in Northern Europe. Inspired by having read the original model description \cite{schelling_dynamic_1971}, where Schelling himself points out that governance and class differences are more relevant to real-life segregation processes than individual action, the author posed a question in the room: Could the Schelling model results be misused to trivialize the role of systemic racism and mask privileges commonly afforded to white communities? The lecturer dismissed this inquiry as irrelevant to the subject of the lecture. According to them, the main interest of the Schelling model was, to quote, \textit{``It is simple and it works.''} The conversation ended there, since none of the other students joined in the discussion. However, it left a deeper imprint on the author: in a quest to find a simple computational explanation for segregation, the violent history of race relations in the United States was being discursively replaced by ``mild preferences for your own kind''. The inference drawn from the Schelling model during this lecture seemed to be that individual choices are the main driver of segregation, and this hypothesis about real-world social dynamics could be computationally proven! How many of the other roughly 70 young students present that day would question such numerical evidence for nuance? How many researchers around the world who study segregation from a computational perspective believe that segregation is largely an outcome of individual choices? And what about the masses? 

The Schelling model is a neat example of how simple rules lead to complex and counter-intuitive patterns, but beneath its surface of mathematical rigour, systemic injustice is disguised as individual choice \cite{porter_trust_1996}. If we want our models to serve society, we need to scrutinize the sociopolitical implications of their assumptions \cite{cottineau_growing_2015, saltelli_five_2020}.

\subsubsection*{Modeling as a Political Process}

Modeling can be understood as a process of abstraction, which necessarily comes with simplifications. As humans, we build mental models on a daily basis; they are key to our interactions with our environment. The task of weekly meal prep, for instance, requires a great deal of algorithmic thinking. However, there are also dangerous examples of abstraction in our everyday lives: the emergence of racist beliefs can be understood as a faulty modeling process, where false premises, self-confirmation biases, and cherry-picking come together to produce dangerous outcomes \cite{oneil_weapons_2017, corbin_terrorists_2017}. In the age of Big Data, the decision-making power granted to computational models is awe-inspiring. Their authority often goes unquestioned, and is coupled with highly detrimental outcomes for marginalized groups: Think of the systemic gender bias in medical treatment trials \cite{perez_caroline_criado_invisible_2019}, or the exclusion of non-white communities in the realm of internet search engines \cite{noble_safiya_umoja_algorithms_2018}, or facial recognition tools failing to recognize Black women as human \cite{buolamwini_gender_2018}. Thus, our modeling decisions have real-life consequences and frequently, these consequences are not just a by-product, but the explicit \textit{goal} when modeling is used for policy and decision-making.

Abstraction, which is the key premise for any modeling process, can both enlighten and obscure \cite{levins_strategies_2007}. Each attempt to model a real-life process is based on decision-making about which elements to leave out, and which to include. A conscious decision to exclude a relevant element from the model, thus changing the model outcomes, can be understood as a strategy that is \textit{actively obfuscating} \cite{levins_strategies_2007}. It is therefore imperative to ask of every model: Where is the rest of the world? For models of segregation, the decision to include or exclude certain elements amounts to an implicit political statement about their perceived relevance. For example, a recent study \cite{liao_uneven_2024} on segregation in a Swedish city frames segregation as a ``critical problem'', and suggests to distinguish between three groups: ``Swedish'', ``Western'', and ``Non-Western''\footnote{We use all three terms in quotation marks here, to visibilize the critiques of the framework from which they are derived \cite{van_schie_datafication_2022}.}. The study quantifies urban segregation using an index that considers relative percentages of ``Swedish'' vs.~``Non-Western'' residents. As a consequence, the corresponding mathematical expression classifies an area as more segregated when more ``Non-Western'' people live there, but as \textit{less} segregated when more ``Western'' people live there (keeping the numbers of the other two groups equal, respectively). Thus, the two political sentiments codified here are: segregation occurs when ``Swedish'' people and ``Western foreigners'' alike are outnumbered by ``Non-Western'' foreigners -- and desegregation can be brought about by increasing the number of ``Western'' foreigners in an area.

\subsubsection*{Schelling's Sociopolitical Origin Story}

Let us apply our suggestion to examine a model's sociopolitical implications by recalling Schelling's original work, where he pointed out that individual choice is \textit{not the most important} driver for segregation. Notwithstanding, his model includes individual choice as the \textit{only relevant} parameter. Schelling clearly articulated his decision to include individual preferences while excluding governmental action and class differences. Schelling, no less, clearly articulated that his \textit{``...~ultimate concern, of course, is segregation by color in the United States’’} \cite{schelling_models_1969}. It is therefore particularly relevant to consider the sociopolitical context of the model. In the late 1960s, when the Schelling model was conceived, racial segregation policies were still legal practice in the US, and had become a prominent political issue \cite{wilson_truly_1987, massey_american_1993}. Residential segregation in 20th century US cities was driven by both individual and systemic racism. However, as Rothstein \cite{rothstein_color_2017} argues, the role of individual racist attitudes has often been unduly amplified, overshadowing other crucial factors such as direct government actions. ``Individual preferences of the inhabitants'' were frequently cited by state actors as a pretext for \textit{justifying} segregationist policies. Throughout the 20th century, government officials and policymakers invoked ``local attitudes'', ``historically established neighborhoods'', and ``desires of the groups concerned'' to uphold segregationist practices within state-sponsored housing site selection and allocation, zoning rules, real estate appraisal, and housing loans \cite{weaver_negro_1948, jackson_crabgrass_1987, johnson_second_1996, rothstein_color_2017}. There is a remarkably clear parallel between these policymakers, who used ``individual preferences'' to implement, maintain, and uphold urban racial segregation, and Schelling, who used ``individual preferences'' as the sole parameter in his model. 

\subsubsection*{Individualistic Bias}

Whether this parallel was intentional is irrelevant to the model's history of reverence. The overemphasis on individual behaviour, also known as individualistic bias \cite{squires_demobilization_2007}, has heavily propagated into mainstream computational approaches to understanding and alleviating segregation at large. The unwavering focus on the individual obscures relations of structural inequality in our society and leaves us unable to imagine solutions outside of existing economic structures. In this essay, we explore how individualistic bias serves to downplay and obscure elements of structural oppression. We argue that engaging with a systemic problem at the level of individual behaviour is likely to generate solutions targeted solely at individuals, while failing to produce incentives for systemic change. Bonilla-Silva expresses this notion as a critique of mainstream antiracism approaches \cite{bonilla-silva_racism_2006} that blame a few ``bad apples'' for their aberrant attitudes while failing to call out systemic oppression. We reveal how individualistic bias extends beyond computational modeling and permeates public discourse, policy-making, and education -- spheres in which we relate to each other as humans in public space. We further argue that these issues are exacerbated by recent developments in artificial intelligence and the datafication of governance. In the digital age, doing Mathematics in the Ghetto without ever having experienced the Ghetto has become very easy. Our hope is to shift attention of our models towards explicitly exposing and encoding systemic drivers of segregation, persuading us to critically engage with the sociopolitical and real-world implications of abstract modeling processes, and reflecting on our positionality as computational scientists. With an appeal to our fellow computational scientists, our vision is that this commentary will foster a space for conscious, community-oriented focus on reducing systemic urban inequalities.

\clearpage
\thispagestyle{empty}
\clearpage

\section{Segregation and the City}
\label{sec:inequality}

\subsubsection*{Spatial Expression of Social Inequalities}

Segregation is well-researched. Yet, we lack a clear, concise, and coherent definition of segregation that holds across all disciplines of research or within the broader public discourse. Several reviews across various geographies and intellectual cohorts -- sociology \cite{wacquant_urban_2008, krysan_cycle_2017}, social and economic geography \cite{harvey_social_1973, musterd_handbook_2020}, law \cite{rothstein_color_2017}, history \cite{nightingale_segregation_2012}, urban planning \cite{espino_building_2015, gans_social_2018}, neighbourhood effects \cite{van_ham_neighbourhood_2012} -- offer differently nuanced framings of the physical separation of people based on race, gender, religion, and other characteristics. One way to formulate a common denominator between these schools of thought is understanding segregation as the \textit{spatial manifestation of social inequality} \cite{blokland_outside_2008}, which is also the interpretation of segregation that we assume for this essay. 

To illustrate the implications of our definition, let us contrast it with two dictionary entries on the term \textit{segregation}. According to Merriam-Webster, segregation is \textit{``the separation or isolation of a race, class, or ethnic group by enforced or voluntary residence in a restricted area, by barriers to social intercourse, by separate educational facilities, or by other discriminatory means''} \cite{merriam-webster_definition_2024}. Cambridge Dictionary suggests that it is \textit{``the policy of keeping one group of people apart from another and treating them differently, especially because of race, sex, or religion''} \cite{cambridge_dictionary_segregation_2024}. When taken to their logical extreme within the social category of race, both these definitions align with the International Criminal Court's definition of apartheid, which, as defined in the Rome Statute, \textit{``means inhumane acts (...) committed in the context of an institutionalized regime of systematic oppression and domination by one racial group over any other racial groups and committed with the intention of maintaining that regime''} \cite{international_criminal_court_rome_1998}. Notably, the crucial difference between the definitions of apartheid vs.~segregation lies in the terms' legal standing. Apartheid, while as topical as ever in 2025 \cite{williams_apartheid_2024, hajir_academia_2025}, is by definition an illegal practice. In contrast, our collective acceptance of the prevalent capitalist economic system as \textit{legal practice} \cite{lipman_segregation_2018} produces segregated societies within the realm of lawfulness. This legal yet problematic idea of segregation is our bone of contention. 

What all of the above definitions have in common, however, is their focus on an individual's belonging to a specific social group by one singled out criterion (e.g., race), which obfuscates all other potentially intersecting factors contributing to that individual's lived experience of segregation (e.g., wealth, health, gender, social context, and exposure to other mechanisms of exploitation and exclusion). Social inequalities, in contrast, are multidimensional \cite{nelson_conceptualizing_2024}. Take, for example, a Black person living in the Netherlands. Our wish is to go beyond the fact that this person is classified as ``Non-Western'' by the census, and in the course of their lifespan, will be unwittingly used as a datapoint for numerous computational models of ethnic segregation in Dutch cities. Apart from being Black, this person is a woman, a parent, a minimum wage survivor, a resident of a flood-prone area in modern day Limburg, and a community leader in her local church. We wish to see this person as a whole, and ask: how do social inequalities manifest spatially in her city, how does she navigate them, and what could -- and should -- we as a society do about these inequalities?

\subsubsection*{The Segregated City}

Why do we focus on cities? Urban areas as an object of study have fascinated humans for thousands of years \cite{vitruvius_vitruvius_1999, jacobs_death_1961}. Contemporary cities can provide the necessary environment for community and solidarity. In densified urban environments, low-carbon lifestyles can flourish, and there is hope that creative ways out of the metacrisis \cite{hagens_many_2024} could emerge from the inspiration and innovation of buzzing urbanity \cite{west_scale_2017}. Granted, these positive attributes are sometimes unjustly romanticized -- but cities undoubtedly bear a potential as hubs of social change. However, inspite of this potential, many contemporary cities contain more doom than gloom, as cities are also the places where oppression and exploitation is concentrated, and where global inequalities are shaped and (re)produced. In the words of Shoshana Zuboff, the city of today is a ``petri dish for the reality business of surveillance capitalism'' \cite{zuboff_age_2018}. Independently of whether we focus on hope or tragedies, today's world is highly urbanized \cite{united_nations_statistics_division_sdg11_2024}, and what happens in cities matters \cite{nijman_urban_2020}.

Segregation, when it is considered as the spatial expression of social inequalities, can materialise at vastly different scales: think of the less industrialized countries that bear disproportionate externalities of the climate crisis \cite{wallimann-helmer_ethical_2019}; or of the gender bias in taking up space at the playground \cite{reimers_physical_2018}; or riots fueled by populism in regions that have been left behind altogether \cite{duncan_local_2024}. While these are relevant to people's lived experience, our narrative and analysis weaves through neighbourhoods and cities, where the majority of segregation literature and public discourse directs its attention, and where reduction of inequalities is an actionable goal.

\subsubsection*{Confronting the Moral Dilemmas of Inequality}

In 2025, the call to reduce inequalities -- United Nations Sustainable Development Goal 10 \cite{un_department_of_economic_and_social_affairs_goal_2024} -- has become a commonplace directive. This focus carries profound implications, both for understanding urban segregation and for driving social change through research, policy, or direct action. For decades, if not centuries, political movements have worked against ill-conceived perceptions of inequality as ``natural'' or ``socially beneficial'' \cite{house_culminating_2019, kallis_limits_2019}. Yet, as the discourse on inequality has become generalized, critical engagement with its deeper philosophical roots and moral implications has often been lost \cite{dixon_investigating_2005}. At its core, addressing societal inequalities is a question of moral philosophy \cite{coleman_inequality_1974}. Are inequalities an inevitable feature of human societies, as conflict theory suggests \cite{simon_conflict_2016}? What constitutes an acceptable -- or even desirable -- level of inequality? Can an ``optimal'' inequality threshold be defined, and by what metrics? Is an inequality-free utopia achievable or even worth pursuing? While this essay does not aim to give prescriptive answers to these questions, we want to address how researchers engage with these questions in the domain of urban segregation and inequality at large, particularly through computational frameworks. 

To more easily group together similar implicit stances of moral philosophy within inequality research, here we propose the term \textit{Engineering-Economy Complex (EEC)} to describe solution-oriented computational research approaches grounded in engineering paradigms and neoclassical economics. The EEC includes, for example, agent-based modeling based on the assumption of rational self-interest, or data science studies whose quantification of inequalities is produced as evidence for policymakers.

Now, let us contrast two epistemological approaches within inequality research: sociology vs.~the EEC. Sociology broadly agrees on inequality's harms, but prioritizes descriptive critique over actionable solutions \cite{house_culminating_2019, diprete_relevance_2021}. Meanwhile, the EEC recognizes inequality as problematic, but lacks consensus on urgency. The EEC operates in \textit{prescriptive mode}, optimizing solutions within predefined system boundaries but rarely interrogating root causes of inequality, and lacking the necessary imagination to come up with an alternative solution space. As EEC members ourselves, we acknowledge our disciplines' limitations. Sociology, conversely, struggles to translate structural critiques into feasible interventions. Both fields risk stagnation without deeper engagement with systemic drivers that ultimately could translate into systemic solutions.

Documenting and describing societal inequalities is a crucial first step in mobilizing collective action to address their causes and consequences. Immediate symptomatic relief, such as welfare provision for the disadvantaged, represents a tangible next step where both the EEC and sociology make huge strides. However, stopping at symptomatic relief risks perpetuating the status quo. To achieve meaningful change, we must engage with the root causes of inequality -- a step the EEC often fails to take. Our EEC is trapped in a ``solution space'' that assumes the boundaries of our current capitalist system as a given. It asks, ``How do we tackle inequalities?'' without questioning the systems that produce them. We argue that addressing root causes is essential for any real change, requiring both conceptual engagement and understanding of individual positionality. For many, including several other academic disciplines and EEC members like ourselves, this means critically examining how we benefit from existing inequalities \cite{schwalbe_elements_2000, simon_conflict_2016}. Such introspection is vital for understanding the EEC's current predicament and breaking free from its constraints.

A recurring issue in inequality discourse is the misalignment between philosophical positions of the stated future goal of eradicating inequalities and the proposed pathways to achieve it, often rooted in individualistic meritocracy or individual welfare. This is the central predicament of the EEC. Not all people genuinely believe that all inequalities can -- or should -- be eliminated \cite{simon_conflict_2016}. The deeper issue is not the validity of this philosophical stance (independently of whether or not we agree with it), but its implicit and unconscious character, as often seen within the EEC. As Victoria Reyes \cite{reyes_academic_2022} points out, this predicament can be better understood when considering that the academy is overwhelmingly white, wealthy, male, and Western; as a result, the institution globally lacks the collective experience of oppression. This lack of experience feeds into the misalignment described above, producing an ideological lock-in. People who work towards inequality reduction, while ignorant of their own disbelief in the feasibility of this goal, heavily constrain the solution space in which the EEC operates.

A poignant example is the moral dilemma faced by parents choosing a primary school for their child in a city with multiple inequalities. In a society where school trajectories significantly shape future life outcomes, even progressive parents may opt for schools with the least accumulated disadvantage, inadvertently perpetuating spatial inequalities. This decision, while understandable, highlights the need to change the boundary conditions of the system we live in rather than the decision-making process itself.

Similarly, and speaking more broadly, as members of the EEC concerned with urban inequalities, our aim should not be to optimize within the existing system but to question the system itself. Solidarity and incremental reforms can only go so far under capitalism. To truly address inequality, we must confront the structural conditions that sustain it, challenging both our intellectual and modeling frameworks, and our personal complicity in the status quo. Toward this goal, below we review the most common perspectives on the \textit{causes} of segregation and then scrutinize the positive \textit{consequences} of segregation to better understand the mechanisms underlying its perpetuation. 

\section{Causes}
\label{sec:causes}

To understand contemporary explanations for segregation from an urban perspective, we must visit the Chicago of 1920s, where the Chicago School of Sociology made several observations about urban communites in the United States and formulated the spatial assimilation theory. This theory, one of the most enduring theories of urban segregation, articulates that residential segregation between two different groups is a function of their socioeconomic and cultural differences \cite{park_urban_1925, park_human_1928}. In his 1920s essay, Park argues that recently arrived immigrants, in rapidly growing North American cities, are not yet familiar with the host country's culture and lack the financial means to settle in more desirable neighborhoods. These two factors together explain the immigrants' initial segregation. Over time, however, the \textit{``keener, the more energetic, and the more ambitious''} \cite{park_urban_1925} individuals will enhance their cultural assimilation and socioeconomic status. Consequently, they will disperse across the city in search of more desirable and expensive neighbourhoods, leading to their integration \cite{logan_growth_1978, massey_ethnic_1985, alba_minority_1993}.

%Inequality drives assimilation
According to spatial assimilation theory, the most salient attribute of the world is social and spatial inequality. Socially, people of different origins also differ in their distribution of socioeconomic resources. Spatially, residential locations differ by land value and desirability. The theory posits a direct proportionality principle for neighbourhood attainment: the higher an individual's socioeconomic capital, the higher the number of desirable residential locations available to them. For spatial inequality to drive spatial assimilation, the implicit assumption is that individuals will aim to improve their lives by moving into more expensive and desirable neighborhoods. Thus, individuals are incentivized to increase their socioeconomic capital, enabling them to make more expensive residential choices -- ultimately moving into (in the US context) predominantly white neighborhoods and vacating their previous residences for newcomers. Here assimilation into the host society is implicitly assumed to be a positive goal and the desirable end state. The theory aligns with a range of subsequent studies focussing on the role of socioeconomic inequalities, particularly wealth inequalities, as the main driver of residential segregation \cite{crowder_wealth_2006} and assimilation over time.

% Problematic adoption of the theory
Park's essay on spatial assimilation \cite{park_urban_1925} is significant for understanding how immigrants and marginalised groups navigate urban spaces, and remains influential in contemporary research \cite{pais_metropolitan_2012, alba_assimilation_2014, andersen_spatial_2016}. As anecdotes go, his observations are accurate for many individual life paths, offering legitimacy to the role of merit in upward social mobility. However, the theory has been challenged for generalising a highly specific context (i.e., white and Asian immigrants arriving in the US at the beginning of the 20th century) and for failing to accurately explain persistent segregation patterns across varying cultural and socioeconomic contexts \cite{charles_dynamics_2003, krysan_cycle_2017}. Ultimately, the theory could be easily misinterpreted when used as a single analytical tool, as it places the entire responsibility of life quality improvement on vulnerable communities themselves, and suggests that only individual effort and ambition can help escape conditions of poverty and deprivation. The theory does not question inequality-producing features of the system -- under which a silent majority, constructed as blameworthy, will always be left behind \cite{houlden_left_2024} by definition -- and focuses solely on individual agency as the driver of change for individual circumstances. 

\subsubsection*{Who can pull themselves up by their bootstraps?}

The limitations of spatial assimilation theory reveal some of the true scandals of our contemporary world. Any theory that acknowledges inequality as a driver of segregation, yet portrays issues of socioeconomic inequality and the lack of good-quality affordable housing as intrinsic and unchangeable features of the system, is reinforcing the Malthusian myth \cite{kallis_limits_2019} that inequality is a ``natural'' phenomenon. Taken in isolation, such theories are suggestive of an individual who has elected or inherited a government that does not consider it a moral responsibility to provide for the people who elected it. Such theories highlight individual effort, affirming the belief that hard work can lead to escaping poverty and vulnerability. While it is widely accepted that market-based economics and political systems play a role in segregation \cite{wilson_political_2008}, theories that focus on socioeconomic differences alone fail to fully explain many of the characteristic patterns of continuing poverty and deprivation \cite{krysan_cycle_2017}. 

Both segregation and socioeconomic inequalities are deeply rooted and temporally persistent in space, making it challenging to disentangle the causal mechanisms \cite{nijman_urban_2020}: segregation can also cause socioeconomic inequalities to arise and reproduce. Place stratification theory, introduced in response to spatial assimilation theory, provides more nuance to this entanglement by describing spatial differentiation \textit{``not only as the population's natural selective response to its habitat but also as a means of organizing inequality''} \cite{logan_growth_1978}. This theory suggests that the stratification of place should be considered an additional dimension of social stratification \cite{davis_principles_1945, hauhart_davismoore_2003, massey_categorically_2007, tilly_durable_2009, peoples_comparative_2012,massey_residential_2016, massey_still_2020}. Different places have differently distributed resources, and according to place stratification theory, these differences are inherently political. Individuals and institutions in advantaged places use their superior resources \cite{logan_growth_1978} to maintain or further improve their position of privilege relative to other places. Thus, inequality both \textit{drives} and \textit{is an outcome of} spatial differentiation maintained and reproduced through political power.

William J.~Wilson's seminal 1987 book ``The Truly Disadvantaged'' \cite{wilson_truly_1987} shifted some focus away from individual agency by illustrating how living in a deprived neighbourhood affects individual life and health outcomes. Although not everyone in Wilson's neighbourhood will ultimately escape their poverty-ridden future (just as the American dream is not attainable for everyone), broader public perception often holds the view that it is an underprivileged individual's own volition to live in poverty, remain jobless, and occasionally lead a life of crime. The academic version suggests that segregation patterns could be explained, at least in part, by individuals' choices and preferences. This idea forms the basis of the well-established stream of research on residential preferences, or choosing where to live.

\subsubsection*{Who can choose where to live?}

Residential preferences refer to an individual's choices and desires for selecting a place to live. These preferences can encompass various aspects, such as preferred size of the dwelling or the proximity to public transport. In the context of segregation research, however, the most frequently studied aspect is that of preferences for specific racial/ethnic neighbourhood compositions, commonly analyzed by methods such as large-scale statistical analysis of census data, stated and revealed preference surveys, and qualitative interviews \cite{farley_chocolate_1978, clark_residential_1991, ihlanfeldt_black_2002, charles_dynamics_2003, adelman_roles_2005, bruch_neighborhood_2006, fossett_ethnic_2006, ibraimovic_changes_2017, krysan_diversity_2017, bruch_choice_2019, liebe_maximizing_2023}. In many cases, these studies embrace the virtue of \textit{tolerance} as an ideological category \cite{zizek_tolerance_2008}, assigning a presumed numerical value to the preferred proportion of another race or ethnicity that one can ``tolerate'' in their neighborhood, which is then interpreted as their willingness to desegregate. It is particularly noteworthy that this stream of segregation research in turn has come to form the theoretical basis for highly influential quantitative approaches such as agent-based modeling, which we will revisit in depth slightly later in the essay.

Racial residential preference studies gained renewed attention in the US in the 1990s, alongside public and political discourse on racial and religious differences, as researchers sought to understand why segregation persists despite government interventions like the Fair Housing Act of 1968 \cite{massey_american_1993, thernstrom_stephan_america_1997}. The self-segregation hypothesis \cite{ihlanfeldt_black_2002} suggests that less privileged groups, such as Black people in the US, purposefully choose to self-segregate \cite{thernstrom_stephan_america_1997, patterson_ordeal_1998}. While this hypothesis lacks broad scholarly support \cite{defina_african_2007}, it remains influential in two ways. First, studies analysing how individual preferences impact segregation outcomes often seek to test this hypothesis based on real-life data; and second, the hypothesis is echoed in public opinion and by bad actors in politics, and is used as a common argument against integration policies \cite{crozier_trouble_2008, naber_zu_2016, cimbidhi_loi_2024, henley_anti-immigration_2024}. This line of thinking implies that desegregation policies are doomed to fail and resources should be directed elsewhere. 

% preference and racism
There are other naive ways in which lines have been blurred between residential preferences and overt racism. For instance, one study on examining white people's racial residential preferences \cite{ihlanfeldt_whites_2004} purposefully disregards the distinction between a ``preference for racial homogeneity'' and a ``distaste for living among Black people'' -- both of which are forms of systemic oppression through racial discrimination. In constrast, as has been repeatedly shown in the US context \cite{kelly_practical_2004, krysan_residential_2002, krysan_does_2009}, Black communities' desire for mixed neighbourhoods is shaped by their experiences with racism and hostility. Recent studies on immigrants in European countries show similar findings \cite{andersen_spatial_2010, doff_residential_2011, ibraimovic_changes_2017, liebe_maximizing_2023}. While these studies may differ in design and approach, they consistently reveal that more privileged groups are ``less tolerant'' (or more racist) in their neighborhood preferences, while less privileged groups tend to prefer more mixed neighborhoods due to their specific urban realities and experiences with forms of discrimination or othering.

\subsubsection*{Who is afraid of the big bad racist?}

Discrimination, particularly racial discrimination in the housing market and mortgage industry, has been extensively studied \cite{massey_use_2001, ross_color_2002, glikman_ethnic_2012, ewens_statistical_2014, andersen_ethnic_2019,roscigno_complexities_2009, anacker_analysing_2011, rugh_race_2015}. While conclusions vary, most sociological research agrees that discrimination against marginalised groups plays an influential role in segregation outcomes. This is particularly evident in places with a long history of colonial violence and forced segregation, such as the Black communities in the US \cite{quillian_racial_2020} or the Muslim communities of Delhi, India \cite{jamil_capitalist_2014}. However, studies using quantitative methods to isolate the effects of discrimination from other factors, such as socioeconomic status or personal preferences, often find only a weak correlation between discrimination and segregation \cite{galster_assessing_1988, south_housing_1998, turner_housing_2013}. Krysan et al.~(2017) caution that these weak correlations should not be interpreted as negligible effects of discrimination. Instead, they argue that the methods used may underestimate the true impact: \textit{``Testing the magnitude of racial differences in the treatment of individuals responding to ads for the same unit is likely to dramatically underestimate the overall effect of discriminatory forces on segregation''}~\cite{krysan_cycle_2017}. 

When discussing segregation, we frequently use language that inherently divides communities. Our data, models, and questions are framed around the ideology of tolerating, rather than embracing, differences. While it is crucial to acknowledge discrimination, an excessive focus on individual discriminatory attitudes as the primary cause of segregation is not helpful for producing evidence that has the power to impact policymaking \cite{blokland_outside_2008}. This approach overlooks the systemic nature of racism, which is obscured when racism is seen solely as an individual mindset, and when emerging forms of institutionalised racism are misunderstood as a sum of individual racist stances \cite{sarbo_rassismus_2022}. In the words of Bonilla-Silva \cite{bonilla-silva_racism_2006}, \textit{``Systems of domination of any kind (...) do not depend for their reproduction on the actions of ``bad actors'', that is, on the behavior of the sexists, racists, elitists, and homophobes. They remain in place fundamentally because of the actions and inactions of many good, nice people. Systemic racism is then maintained not by the actions of the ``racists'', but by the passive, habitual, and mostly neutral behavior of the majority of Whites (neutrality in the face of systemic inequality amounts to supporting the status quo).''} 

Much like spatial assimilation theory, residential preference studies mostly implicitly assume that individuals make choices freely and rationally, borrowing from neoclassical economic theory. However, human behavior is far more complex than self-interested decision-making with perfect information. Yet, most studies on residential mobility and segregation traditionally focus on the role of economics, preferences, or discrimination, trying to assess their relative importance \cite{krysan_cycle_2017}. These studies frequently overlook processes of decision-making and housing search. When housing search is considered, it is usually modeled through a traditional economic lens, assuming perfect, uniform, and unbiased (``colourless'') knowledge. 

\subsubsection*{What makes housing search a cultural practice?}

To offer a more nuanced perspective, Krysan et al.~(2017) develop the social structural sorting theory, suggesting to consider an individual's lived experience, daily activities, and social perceptions -- shaped by their neighborhood -- as integral to decision-making processes. This perspective articulates that people's choices are shaped by their positionality and lived experiences, both of which are deeply influenced by systemic inequalities they face \cite{krysan_cycle_2017}. Additionally, multiple other psychological and socio-economic processes play a significant role in decision-making. This, in turn, affects residential mobility outcomes, and can contribute to segregation patterns. Within progressive economic theory, other schools of pluralist economic thinking, including behavioral economics, better accommodate this reality \cite{brand-correa_economics_2022}. Social structuring sorting theory, perhaps influenced by progressive economic thought, has sparked further research integrating cognitive science into housing search studies \cite{bruch_choice_2019, deluca_not_2020}. 

\subsubsection*{But where is the macro-level scrutiny?}

In summary, most research approaches to \textit{causes} of segregation focus on individual-level explanations. They scrutinize individual agency, behaviour, choice, preference, experience, or racist stances, to explain segregation's emergence and persistence. This aligns with viewing segregation as a macro-level outcome arising from micro-level interactions \cite{schelling_micromotives_1978, wiley_micro-macro_1988, stadtfeld_micro-macro_2018}, a framework central to computational approaches. These models simulate micro-level dynamics to reproduce macro-level segregation, mirroring Schelling's classic ``emergent property'' paradigm. Yet this lens captures only half the story. Causality flows bidirectionally: macro-level structures (e.g., capitalism’s growth-inequality cycle) predetermine the solution space within which individuals operate. Systemic inequalities under capitalism constrain choices long before individuals decide where to live. If we aim to disrupt the status quo of urban segregation, we must scrutinize these macro-level forces with the same rigor that has historically been applied to micro-level behavior. To focus solely on individual agency is to ignore the structural oppression that shapes -- and dictates -- the choices we study.

\section{Consequences}
\label{sec:consequences}

Discussions on the consequences of spatial segregation commonly and inevitably evoke the suffering of the marginalized. According to the Engineering-Economy Complex (EEC), spatial segregation is implicitly understood as a kind of societal illness which must be cured \cite{anderson_imperative_2013, shelby_integration_2014}, and the EEC sees itself as a space of expertise that has the answers. In the book ``The Will to Improve'' \cite{li_will_2007}, anthropologist Tania Li describes how experts draw boxes around their craft and only view problems within those boxes, disregarding what social and political challenges lay outside of them. The EEC has good intentions: we want to support marginalised individuals in accessing resources, such as transport, housing, and energy. However, we rarely question how such inequalities came to be in the first place. The most recurring stance within the EEC is to dissociate our expertise from the object, proclaiming that segregation is a characteristic deficiency of those \textit{negatively} affected by it and they must be helped through technocratic solutions \cite{li_will_2007}.

\subsection*{The Benefits of Living a Segregated Life}

Certainly, segregation has tremendous negative consequences for those already oppressed and vulnerable \cite{wacquant_urban_2008, krysan_cycle_2017, kirk_making_2019}. However, by framing it as a problem confined to the experiences of the poor, the vulnerable, and the marginalised, we render larger structures and systems of oppression invisible. This narrow focus directs attention toward ``fixing'' marginalised lives, avoiding dismantling the systems that simultaneously perpetuate oppression and sustain the privilege and wealth of others. Such an approach is fundamentally inadequate for achieving systemic change. If our goal is to meaningfully alter the trajectory of inequality, it is imperative that we move beyond mere acknowledgment of the suffering endured by the disadvantaged. We must explicitly confront the agency and accountability of those who derive benefit from dominant systems of oppression, including ourselves.

\begin{figure}[ht]
\centering
\includegraphics[width=0.85\textwidth]{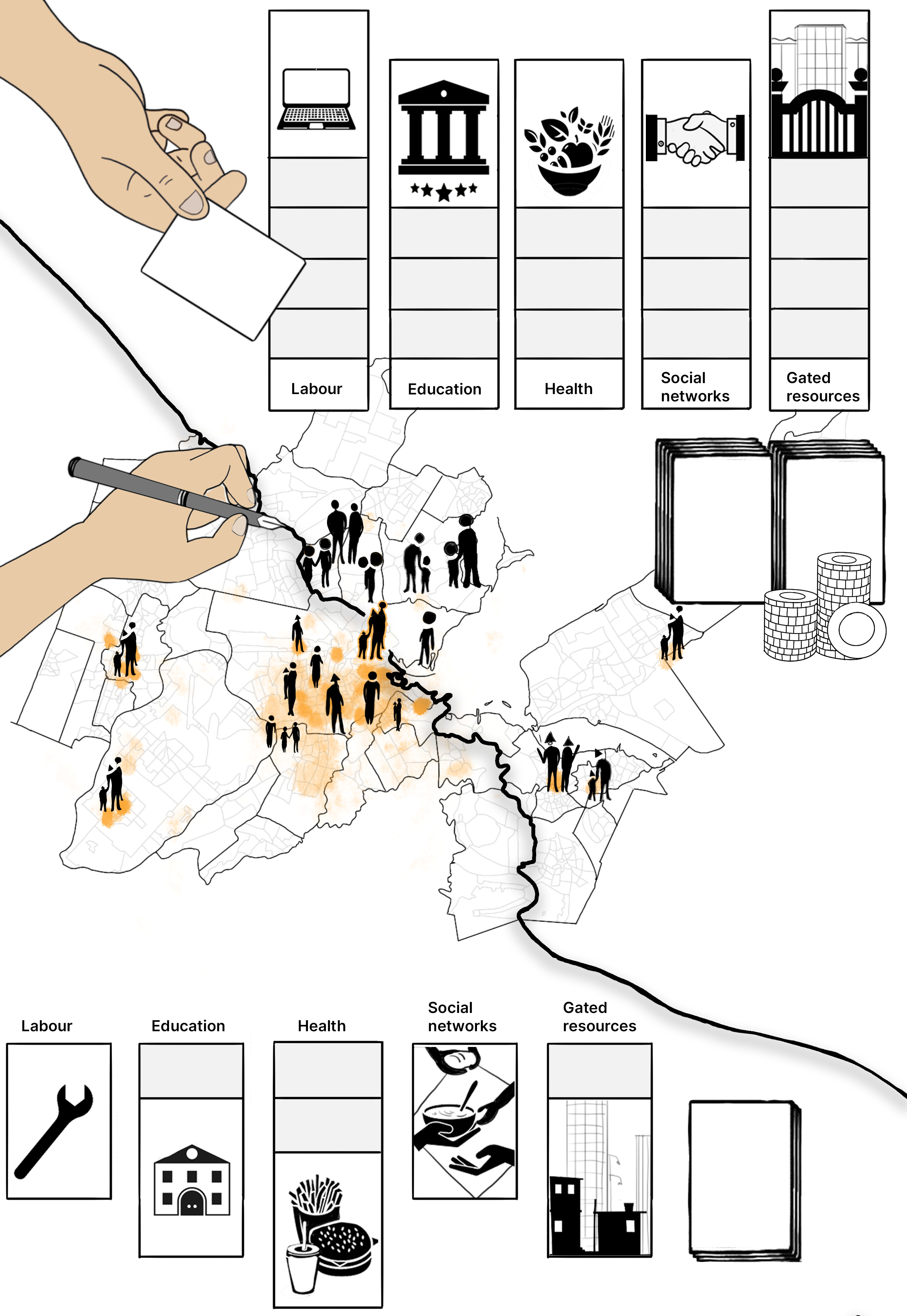}
\caption{The benefits of living a segregated life: spatial inequalities skew the resource distribution along multiple axes, such as labor, education, health, social networks, and gated resources. These axes are differently configured in advantaged (top) vs.~disadvantaged (bottom) neighbourhoods. People who hold power to mitigate these inequalities are not always conscious of the advantages they draw from them. Meanwhile, gaining consciousness of the ways in which the benefits of segregation are distributed in the city and in society at large is a necessary step towards systemic change.}
\label{fig:causesandconsequences}
\end{figure}

To frame segregation solely as ``bad for the poor'' is to neglect its corollary: that segregation is, in many ways, ''good'' for middle and high income groups. These two perspectives are deeply linked, representing two sides of the same coin, yet the solutions they imply diverge significantly depending on where we direct our focus (see Figure~\ref{fig:causesandconsequences}). A discourse centered on the plight of the marginalised, risks devolving into paternalistic pity, leaving the structures of privilege (and oppression) unchallenged. To disrupt this cycle, let us shift our inquiry and ask: \textit{cui bono?} -- who benefits? This question compels us to critically examine the consequences of segregation not only for the disadvantaged but also for the overadvantaged. In doing so, our hope is to unravel the mechanisms through which segregation sustains inequality and interrogate the vested interests that resist its dismantling. Only then can we collectively envision and pursue (by computation if necessary) transformative solutions that address the root causes of systemic oppression and not just its symptoms.

\subsection*{Rich get Richer! But how?}

First, it is essential to recognise the clear socioeconomic advantages afforded to individuals living in more affluent areas. Empirical evidence consistently demonstrates that the wealthier the neighbourhood in which one grows up, the greater the likelihood of financial success later in life. This stratified advantage is not an individual phenomenon -- it is perpetuated intergenerationally, threading inequality into the fabric of our society. This pattern has been robustly documented in sociological theory and supported by quantitative longitudinal analyses \cite{ananat_wrong_2011, beaulieu_benefits_2011, lipsitz_how_2011, krysan_cycle_2017, andersson_segregation_2018, lipman_segregation_2018}. 

In a world where resources are finite, comparative advantage becomes a critical determinant of opportunity, and spatial inequality systematically benefits certain areas at the direct expense of others. For example, the scarcity of job opportunities in disadvantaged neighbourhoods \cite{korsu_job_2010, krysan_cycle_2017, bastiaanssen_does_2022} cannot be understood in isolation; it is inherently \textit{relational}, as the absence of employment in these areas is a direct consequence of the clustering of jobs \textit{elsewhere} \cite{beaulieu_benefits_2011}. This dynamic extends beyond employment to encompass other critical resources that enhance urban life quality, such as access to well-maintained built environments through private and unified land ownership, quality education facilities, reliable public transport, green spaces, and retail and health amenities. Crucially, simply adding more of a certain resource in an underprivileged neighbourhood without considering the broader context of its situation can precipitate gentrification processes and thus have the opposite effect, further entrenching and worsening spatial inequality by displacing existing communities  \cite{tierney_gentrification_2015, uitermark_gentrification_2007, hoffman_bike_2016}.

\subsubsection*{Gated Resources}

Unequal spatial distribution of vital urban resources incentivizes a process that Charles Tilly \cite{tilly_durable_2009} has titled \textit{social closure}: the practices, whether conscious or not, by which dominant groups secure exclusive access to resources while systematically restricting access for others \cite{anderson_imperative_2013, shelby_integration_2014}. This phenomenon manifests in various forms across the globe, from the historical patterns of ``white flight'' in the United States \cite{boustan_was_2010, chang_white_2017} to the proliferation of gated communities in Brazil \cite{monteiro_enclaves_2008}, and nepotistic hiring practices in Italy \cite{gagliarducci_politics_2020}. More broadly, the self-segregation of the wealthy -- whether through exclusive neighborhoods, private institutions, or social networks -- exemplifies how resource-controlling groups actively reinforce their privilege \cite{gans_involuntary_2008, blokland_outside_2008}. Crucially, the greater the spatial inequalities, the stronger the incentive for these groups to insulate themselves from others, perpetuating a cycle of exclusion and deprivation \cite{andersen_ethnic_2019}. 

\subsubsection*{Cheap Labour}

Concentrated disadvantage serves an important function for those in positions of power: it guarantees a readily available supply of cheap labor force. In the United States, where race and economic status are deeply intertwined \cite{rugh_race_2015, massey_still_2020}, Pauline Lipman has described this dynamic as \textit{racial capitalism} with a distinct spatial dimension. Here, the existence of low-wage workers -- spatially segregated and easily displaced -- is not merely a byproduct of inequality, but a deliberate mechanism that benefits higher-income groups \cite{lipman_segregation_2018, lipsitz_how_2011}. This pattern is not confined to the US; South and South-East Asia are tragic examples of vastly urbanising lands where economic benefits are locked into gated communites at the expense of, not alongside, exploited populations of undereducated and underpaid workers \cite{spencer_emergent_2011, nahar_lata_thats_2022, roitman_understanding_2020, roy_chaudhuri_normalized_2021}. In these contexts, economic growth and development get concentrated in enclaves of privilege, while the surrounding populations bear the costs of exploitation and exclusion.

\subsubsection*{Education}

Segregation also confers significant educational advantages to groups with greater socioeconomic capital, reinforcing and perpetuating their privileged position. High-quality schools that are concentrated in wealthier areas provide superior educational opportunities, which are further amplified by the additional resources affluent families can mobilise, such as social status, cultural capital, access to private tutoring, informational networks, and internship opportunities. These advantages are critical, as education is a key determinant of life outcomes \cite{leathwood_social_2004}. In systems where school funding and resources are tied to the socioeconomic standing of their surrounding areas, educational inequalities are particularly pronounced, creating a self-reinforcing cycle of privilege and gatekeeping \cite{owens_social_2019, jung_spatial_2014, burger_socio-spatial_2019}.

Those who benefit from such systems also gain from the artificial suppression of competition for university admissions and other opportunities. Exclusionary mechanisms, though sometimes overt -- such as the historical and ongoing systematic exclusion of women from education \cite{avolio_factors_2020, roumell_social_2021, kumar_taliban_2024} -- are otherwise more subtle when operating along socioeconomic lines. These less visible mechanisms, such as unequal access to quality schooling or extracurricular resources, are effective in maintaining barriers for disadvantaged groups. 

\subsubsection*{Health, Wealth, and Life Outcomes}

Additionally, segregation perpetuates unequal health outcomes, creating a clear divide between those who can afford to live in safer, healthier environments and those who cannot. The unequal distribution of health risks, such as exposure to environmental hazards, air and noise pollution, and heavy traffic, systematically benefits wealthier groups who can distance themselves from public health hazards \cite{tehrani_color_2019, yang_racialethnic_2020}. In contrast, disadvantaged communities continue to bear a disproportionate share of negative externalities, leading to poorer health and reduced quality of life \cite{guimaraes_racial_2022, beck_color_2020, marinacci_role_2017}. An illustrative example of this dynamic is the longitudinal impact of highway construction in the US \cite{aiello_urban_2025}. While highways provide convenience to those who can afford to live farther away from them, they simultaneously degrade the health and life outcomes of those living in close proximity, who are disproportionately low-income and minority populations \cite{nall_road_2018}.

Health is unequivocally tied to wealth, which is a significant determinant of longevity. Wealthier individuals not only have greater access to healthcare but also live in environments that promote better health outcomes. In regions with high inequality and in absence of a socially robust healthcare system, access to healthcare is highly uneven, exacerbating differences in life expectancy and overall well-being \cite{beck_color_2020, nelson_conceptualizing_2024}. Additionally, both physical and social mobility play a critical role. Wealthier individuals can more easily relocate to healthier environments or access preventative care, using the expensive car apparatus \cite{mattioli_political_2020}, while those in disadvantaged communities face barriers across different key aspects of their life, such as economic hardship, lack of inexpensive mobility options, poor quality food choices, and deprived built environments, to name just a few. Collectively, these barriers majorly reduce access to preventative healthcare. 

\subsubsection*{Social Networks}
Human beings are social creatures. For us, urban areas bear the promise of social connections and improvement of livelihoods. As sociologists, network scientists, and computational social scientists have demonstrated, who one knows matters in life; a person’s network can be a critical source of support, information, and opportunity, which plays a pivotal role in determining life opportunities and outcomes \cite{espin-noboa_inequality_2022, iniguez_universal_2023, lindh_social_2024, reme_quantifying_2022, schmutte_job_2015, asikainen_cumulative_2020}. Social and spatial segregation deeply affect these social ties. The opportunities shared among peers will have differing value and impact, depending on socioeconomic status \cite{qadeer_ethnic_2006}. For people with fewer resources, networks of care are vital for survival and resilience \cite{small_how_2020, deckard_poor_2022, mijs_how_2024}, while for those with accumulated wealth, networks of support function as tools for preserving status and amplying privilege \cite{twohey_weinsteins_2017, klenk_wie_2021, klenk_warum_2024}. 

In highly unequal societies, upward social mobility is severely constrained \cite{mitnik_social_2016, dribe_urban_2024}, and segregation acts as a gatekeeping mechanism that limits the flow of opportunities through personal networks. Of course, social proximity and physical proximity are not identical, but the spatial dimensions of our lives undeniably shapes our social realities. Living in a wealthier neighbourhood, for instance, provides not only material advantages but also cultural capital -- skills, norms, and connections that make it easier to integrate into privileged circles \cite{otero_space_2022, gerxhani_cultural_2022, nieuwenhuis_does_2020}. In this way, segregation functions as a bootcamp for acquiring the cultural fluency needed to navigate and thrive in elite social networks.

\subsubsection*{Out of Sight, Out of Mind}

Ultimately, segregation serves a moral function by allowing those in privileged positions to remain insulated from the realities of systemic disadvantage. When we are not directly confronted with the concentrated hardships faced by others, it is easier for us to accept -- or outright ignore -- the existence of such inequalities and our complicity in perpetuating them \cite{kelly_practical_2004, mijs_how_2024}. This physical and social distance graciously oils a mindset of individualistic meritocracy, where success is framed as the result of personal effort, and failure a lack thereof. In this narrative, we convince ourselves that we deserve what we have because we ``worked hard'' for it, and that others could achieve the same if only they were equally diligent \cite{mijs_is_2021, lindh_social_2024}.

Sheltering us from the reality that some others work harder for far less, segregation allows us to avoid confrontation with our own privilege and the unjust power relations that sustain it \cite{bonilla-silva_racism_2006, haynes_ghetto_2008, mijs_how_2024}. This avoidance stifles meaningful engagement with radical solutions to inequality, as it is easier to dismiss redistributive policies when the beneficiaries are portrayed as undeserving. In the words of Herbert J.~Gans (1972), ``redistributive alternatives can be made to look quite unattractive if those who will benefit from them most can be described as lazy, spendthrift, dishonest, and promiscuous'' \cite{gans_positive_1972}. This explains the persistent use of loaded terms like ``ghetto'' in public discourse, which conjure images of an undeserving underclass and justify exclusionary policies \cite{seemann_danish_2021}. For example, Denmark’s so-called ``Ghetto laws'' illustrated how such rhetoric can legitimize large-scale social housing commodification under the guise of addressing urban decay, a process that Bjarke Risager describes as racial neoliberalism \cite{risager_territorial_2023}. 

\subsection*{Segregation begets more segregation}

Poverty serves positive functions within society \cite{gans_positive_1972, clark_dark_1989}. This appalling fact persists even in the face of widespread societal agreement that it should be eradicated \cite{kelly_practical_2004}. The same applies to socio-spatial inequalities: our moral compass is in agreement that these should not exist \cite{chang_white_2017}, but as long as those who benefit from spatial inequality are doing well under a capitalist system, there is little incentive for change. The most insidious consequence of spatial inequality, therefore, is its self-perpetuating nature. It creates and reinforces the very conditions that sustain it, persistently engendering incentives for those in privileged positions to maintain the status quo. This cyclical dynamic ensures that spatial inequality remains deeply entrenched, even as moral and ethical arguments against it grow louder. 

\clearpage
\thispagestyle{empty}
\clearpage

\section{Computation}
\label{sec:computation}

Let us recall our goal. To meaningfully address inequality, it is essential that we challenge our intellectual and modeling frameworks, and our complicity in maintaining the status quo. Having established the foundational mechanisms that reproduce segregation and the overlooked positive consequences that help entrench these patterns, we now turn to the role of computation. In this next section, we critically assess how computational approaches across various domains distort the framing of socio-spatial inequalities, reinforcing reductionist, individualistic approaches to address segregation.   

The strategic collection and use of Big Data to advance political projects such as nation-building is hardly a new phenomenon \cite{anderson_imagined_1983}. However, in recent years we have witnessed an unparalleled increase of public interest in AI (Artificial Intelligence)\footnote{or SALAMI (Systematic Approaches to Learning Algorithms and Machine Inferences) \cite{quintarelli_lets_2019}}, spilling over into a wide range of knowledge domains, including urban planning and policy-making. This trend, coupled with the growing availability of vast data sets and an increasing willingness to tolerate invasive personal data collection \cite{thompson_smartphones_2019}, explains why data-driven methodologies are gaining traction in governance and decision-making \cite{kandt_smart_2021}. In the discussion that follows, we contend that the discursive hype around AI and Big Data substantially impacts, and at times distorts and misleads, computational approaches to urban segregation through a phenomenon which we term ``New Essentialism''. We then illustrate how this concept manifests across different computational approaches for analyzing urban segregation, including spatial segregation indices, agent-based modeling, and machine learning-driven forecasting.

\subsection{Data-driven New Essentialism}

% what do we mean by re-ethnitization?
The dominant quantitative approach to (urban residential) segregation follows a two-step logic:
\begin{enumerate}
    \item[] \textbf{Step 1.} Identify a set of distinct social groups.
    \item[] \textbf{Step 2.} Analyze how these groups are distributed across fixed spatial units. 
\end{enumerate}
\noindent For the second step, increasingly sophisticated mathematical models -- such as machine learning (ML) techniques, agent-based models (ABMs), and statistical forecasting -- continue to evolve. However, the first step, categorizing social groups, remains strikingly simplistic. The underlying logic is simple: computational approaches require input data, and for practical purposes, these data must be structured into categories that distinguish between social groups. Since discrete categories are easier to handle than fluid or context-dependent ones, simplification becomes inevitable. For example, answering the question ``What is your nationality?'' lends itself to a neatly defined category of citizenship, which is readily codified. In contrast, responses to ``In which social situations is your sense of belonging evoked?'' are far more complex and resist easy classification \cite{franklin_making_2023}. Identity and sense of belonging are inherently fluid and, arguably, unquantifiable. Yet, for the discernment of social groups as clear-cut input to computational models, discrete categories are preferred \cite{schneider_social_1993}. Within the EEC, even if and when there is an implicit understanding that these categories might not perfectly -- or even adequately -- represent reality, they are still used out of sheer practicality, simply because they allow the models to run.

This issue is particularly evident in the reliance on census data for segregation studies, without an interrogation of how and why census categories are constructed in the first place. Census categories impose formal distinctions between groups, which is an inherently political act. For example, several Northern European countries have introduced the category of ``Non-Western'' in their censuses, effectively sorting migrants and their native descendants into two broad camps: ``Western'' vs.~``Non-Western''. This classification reinforces the misguided assumption that so-called ``Western'' migrants are not only geographically, but also culturally closer to the native population \cite{cbs_person_2024, seemann_danish_2021, andersson_comparative_2018}. The function of this distinction is clear: to establish a socioeconomic hierarchy of belonging \cite{risager_territorial_2023}. 

And so, a bitter contradiction emerges. Scientific advancements now allow us to map our complex genetic heritage and imagine our ancestors' migration trajectories for the price of a DNA kit \cite{mcclain_ignoring_2018}. At the same time, as computational methods for policy-making and research (Step 2) grow increasingly sophisticated, the categorization of mutually exclusive race and ethnicity (Step 1) remains alarmingly reductive -- reverting to rigid, mutually exclusive labels that fail to capture the nuances of social processes and realities. Worse still, feeding these crude categories as variables into a hyper-complex ML model is not just devoid of critical thought; it has dangerous consequences for policymaking. Instead of moving beyond divisive categories, data-driven research and policy-making are, in fact, often reinforcing them. Through the growing opacity and complexity of computational methods, the profound consequences of these choices are being further obscured. This is the essence of Data-Driven New Essentialism: a phenomenon in which the rise of computational approaches not only perpetuates categorical simplifications, but does so under the guise of scientific rigor.

\subsection{Spatial Indices}

% quantitative: spatial indices.
The quest to quantify segregation can be traced back to a pivotal academic debate that started in the 1940s in the US, unfolding against the backdrop of legally enforced racial segregation under Jim Crow laws. Early studies \cite{jahn_measurement_1947, hornseth_note_1947, williams_another_1948} drew upon existing methods from ecology and economics and applied them to US census data, where race was one of the recorded variables. These studies sought to distill urban segregation patterns into a single numerical index, structured around a rigid ``Black vs. non-Black'' dichotomy. The objective was twofold: to compare segregation levels across different cities and to examine correlations between segregation (as expressed by the index) and various socioeconomic factors. Notably, the resurgence of the term ``ghetto'' as a designation for an underprivileged urban area in US-American sociology of the 1960s was, in part, a consequence of this growing preoccupation with \textit{measuring} segregation \cite{haynes_ghetto_2008}.

Today, the debate over how to quantify spatial segregation remains unresolved. Various studies have attempted to refine segregation indices by incorporating measures of multigroup segregation \cite{reardon_measures_2002}, human mobility \cite{candipan_residence_2021, iyer_transport_2023}, place connectivity \cite{roberto_spatial_2018}, and non-parametric or multidimensional spatial distribution metrics \cite{sousa_quantifying_2022, spierenburg_characterizing_2023}. These studies have undoubtedly increased computational sophistication compared to the methods of the 1950s. At the same time, most of these studies are instances of New Essentialism we describe above: despite disclaimers that their methodology could be applied to any population variable, they invariably default to considering race \cite{brown_spatial_2006}, ethnicity \cite{olteanu_multidimensional_2020, georgati_disaggregation_2023}, or non-Westernness \cite{spierenburg_characterizing_2023} as primary analytical focus.

A striking example of New Essentialism institutionalized in government policy-making based on spatial indices is the Leefbaarometer (``Liveability Meter'' in Dutch). The Leefbaarometer is a nationwide policy instrument of the Dutch Ministry of Interior and Kingdom Relations. For over two decades, the Leefbaarometer has been employed to ``periodically measure liveability at a small-scale spatial level, so that it can be determined at an early stage in which neighbourhoods there are worrying developments''\footnote{Translation from Dutch by the authors} \cite{ministerie_van_binnenlandse_zaken_en_koninkrijksrelaties_faq_2024}. This tool computes a so-called ``liveability index'' at a granular resolution of 200x200 meters, derived from a regression model incorporating various indicators based on both stated and revealed preference surveys. Municipalities rely on this index to inform urban policy decisions, making it a cornerstone of Dutch urban governance \cite{uitermark_statistical_2017}. 

The policy implications of this index are not neutral. The Dutch Ministry of Interior and Kingdom Affairs actively encouraged municipalities to use the Leefbaarometer when applying the Act on Extraordinary Measures for Urban Problems. This controversial legislation, initially introduced in Rotterdam in 2006 and subsequently adopted in other cities, effectively allows municipalities to prohibit low-income individuals who have lived in the city for fewer than six years from moving into designated neighborhoods. The stated goal of this policy is to enforce social mixing -- by restricting the rights of the most disadvantaged populations \cite{uitermark_statistical_2017, van_gent_exclusion_2018}. 

Between 2014 and 2020, the Leefbaarometer 2.0 explicitly incorporated ethnicity-based indicators, namely the percentage of all \textit{non-Western} foreigners, further singling out specifically the percentage of residents of Central and Eastern European origin, and the percentage of residents of Moroccan, Surinamese, and Turkish origin, respectively \cite{leidelmeijer_leefbaarometer_2014, leidelmeijer_leefbaarheid_2014}. These ethnicity-based indicators were factored into the model as correlating with a lower quality of life in the respective area.  Proponents of the Leefbaarometer defended this approach by arguing that the model does not actively discriminate, as it merely establishes correlation instead of causation. Their reasoning was that if certain ethnic groups are structurally disadvantaged, then their presence in a neighborhood could reflect preexisting socio-economic conditions rather than create them. From this perspective, they claimed, the Leefbaarometer could actually \textit{help} these groups by identifying areas in need of intervention \cite{van_alpen_dit_2020}. 

This argumentation displays several incoherencies. First, the logic of estimating a spatial variable (life quality in a neighbourhood) via a non-causally correlated variable (ethnicity) to then design policies aimed at improving the former by targeting the latter is not only methodologically flawed but also self-reinforcing. If one strictly adheres to the Dutch government’s definitions of ``Non-Westerners'' and ``life quality'', then removing people of non-Western origin would, by definition, improve neighborhood liveability -- a conclusion that is simultaneously both nonsensical and alarmingly actionable. Second, given that the Leefbaarometer 2.0 partially relied on stated preference surveys -- where residents were explicitly asked whether they felt that living next to a Turkish or Moroccan neighbor affected their quality of life -- the regression model effectively merges material urban infrastructure (such as parks and bike paths) with contemporary Dutch racial anxieties. In doing so, it quantitatively institutionalizes these perceptions into data-driven policy making.

After public protests and parliamentary debates, the most overtly migration-related indicators were removed from the model in its latest iteration, Leefbaarometer 3.0 \cite{mandemakers_leefbaarometer_2021}. The broader issue remains: even as computational methodologies become more complex, they continue to reinforce essentialist categories that obscure their political and social ramifications. Spatial indices that rely on race, ethnicity, or migration-based proxies continue to uphold outdated categorizations, hiding the real discriminatory effects of such policies.

\subsection{Agent-based Modeling}

Agent-based modeling (ABM) is becoming an increasingly popular approach across various disciplines, including computational social sciences. Notably, the model often cited as the first (and most canonical) ABM is Schelling's model of segregation. With very few exceptions to date \cite{pangallo_residential_2019, dignum_data-driven_2024}, nearly all contemporary ABMs of segregation trace their lineage back to Schelling's work -- sometimes explicitly, sometimes implicitly. As a result, many discussions of ABMs and segregation today contain veiled references to Schelling's model -- or, perhaps more accurately, to Sakoda's. 

The first checkerboard-style ABM of social interaction dynamics, preceeding Schelling’s, was created by JM Sakoda, whose experiences as a Japanese-American imprisoned in a World War II detention camp shaped his understanding of social interactions under asymmetric power conditions. Over the course of three years, Sakoda observed the deeply unequal interactions between Japanese-American detainees and the camp's administrative personnel. This inspired him to formalize these interactions mathematically, leading to a model that captured power imbalances within group dynamics. After regaining freedom, he developed this work further in his PhD thesis and subsequent journal publications \cite{sakoda_minidoka_1949, sakoda_checkerboard_1971}. 

Although strikingly similar in methodology, Sakoda's model was more generalized than Schelling's later models \cite{schelling_models_1969, schelling_dynamic_1971}. There is little evidence to suggest that Schelling was aware of Sakoda’s work at the time \cite{hegselmann_thomas_2017, aydinonat_interview_2005}. The model that is now universally known as \textit{the} Schelling model is arguably just one specific instance of Sakoda’s broader conceptualization. The relative invisibility of Sakoda's contributions and, in contrast, the unbridled and near-universal popularity of the Schelling model, is less a reflection of the comparative scientific merit of their work and more an artifact of historical timing, disciplinary bias, and the broader politics of academic recognition \cite{hegselmann_thomas_2017}. 

In his first study on segregation modeling \cite{schelling_models_1969}, Schelling's agents are placed on a simple one-dimensional line. He later extended this framework to two dimensions, leading to what is now recognised as ``the'' Schelling model \cite{schelling_dynamic_1971}. The essence of the Schelling model can be summed up as follows: Each agent belongs to one of two groups. To initiate the model run, each agent is randomly placed on one of the grid cells. There are more grid cells than agents, so some cells are left empty. Each agent has a local neighbourhood, consisting of the 8 surrounding grid cells (queen contiguity). Each group has a predefined ``tolerance'', a threshold dictating the maximum proportion of out-group neighbors an agent can ``tolerate'' before deciding to move. There is only one behaviour rule for all agents: If an agent finds itself surrounded by more out-group members than its tolerance threshold allows, it relocates to a randomly chosen vacant grid cell. The simulation runs until an equilibrium is reached (i.e., when no agents wish to move). For a broad range of parameter values (group size, grid size, and tolerance thresholds), the model invariably converges to a state of pronounced spatial clustering, where large homogenous clusters of same-group agents dominate the grid. 

Although the Schelling model can, in theory, be applied to any binary classification, Schelling made his motivations explicit: ``My ultimate concern is of course segregation by color in the United States'' \cite{schelling_models_1969}. Hence the two constructed groups in this model's origin story are Black and white people. Schelling's key conclusion was that ``the interplay of individual choices, where segregation is concerned, is a complex system with collective results that bear no close relation to individual intent'' \cite{schelling_models_1969}. This premise was later generalized in his 1978 book ``Micromotives and Macrobehaviour'' \cite{schelling_micromotives_1978}. Here, Schelling expands on his ideas, drawing from his training as an economist and particularly linking with Adam Smith's concept of the invisible hand \cite{klein_agentenbasierte_2018} which says that when every group member acts in their own interest (i.e., following ``micromotives'' as a fully rational, self-interested agent), the overall outcome (i.e., the observed ``macrobehavior'' of the group) might not be the most satisfying for any of the group members. The resulting argument was that segregation could emerge as an unintended consequence of individual choices, even in the absence of overtly racist intent.

Over the past two decades, research inspired by the Schelling model has proliferated \cite{ubareviciene_fifty_2024}. There are several discerneable reasons for this trajectory. With the advancement of ICT infrastructure and Big Data, agent-based modeling as an approach has gained a lot of traction. The Schelling model is conceptually very simple both to understand and to reproduce, so it has found its way into research and classrooms alike \cite{complexity_explorer_santa_fe_institute_agent-based_2024}. Moreover, the persistent relevance of urban segregation has kept it at the forefront of scholarly inquiry, and the Schelling model is discursively tightly coupled with this inquiry. However, its mainstream reception has led to a critical distortion: an implicit and sometimes explicit suggestion that the model explains real-world segregation patterns, despite its failure to account for systemic, structural, and historical factors \cite{ihlanfeldt_black_2002, bruch_neighborhood_2006, pancs_schellings_2007, fossett_effects_2009, batty_defining_2021, stepinski_machine-learning_2022, hatna_combining_2015, gambetta_mobility_2023, ubareviciene_fifty_2024}. For instance, Urselman’s 2018 study states \cite{urselmans_schelling_2018}:

\begin{displayquote}
    ``In the context of widespread racial segregation in the US, the \textbf{model could demonstrate} that for segregation to occur on a macro scale, \textbf{no deeply entrenched racism was required}.'' (emphasis by authors) 
\end{displayquote}

\noindent While not logically incorrect, this assertion is a misleading abstraction -- it diverts attention away from the fact that deeply entrenched racism \textit{was} indeed the driving force behind legally enforced segregation in the US. A similar misrepresentation of segregation drivers is found in Liebe et al.~(2023) \cite{liebe_maximizing_2023}:

\begin{displayquote}
    ``[C]hoice modeling research – the modeling of decision processes and outcomes – can help to uncover how individual behaviour varies across social contexts as well as the microfoundations of many macrophenomena. One of the most striking examples of this is residential segregation (Sakoda 1971; Schelling 1971), where subtle behavioural responses at a microlevel are \textbf{well known to have the potential to lead to severe consequences such as ethnic segregation} at the macrolevel.'' (emphasis by authors)
\end{displayquote}

\noindent Once again, the language subtly implies that individual micro-decisions alone can account for large-scale ethnic segregation, without acknowledging the systemic and policy-driven structures that reinforce it. Perhaps the most problematic extrapolation comes from WAV Clark \cite{clark_understanding_2008}: 

\begin{displayquote}
    ``Schelling's analyses of the implications of preference schedules suggested that \textbf{mixed-race residential neighbourhoods are not likely to be stable.}'' (emphasis by authors)
\end{displayquote} 

\noindent This claim misrepresents Schelling’s findings. His model merely demonstrated that, under a specific set of assumptions, where agents operate in a featureless 2D grid world without external influences like a government, complete integration is difficult to sustain. What Schelling's model does not prove is that mixed-race neighborhoods (Black and white people) in real-world societies are inherently unstable. And these misinterpretations have real consequences on policy-making. Clark himself used the Schelling model as an argument in his testimony before the US Supreme Court, claiming that desegregation policies were unlikely to succeed \cite{us_supreme_court_freeman_1992, stubbs_freeman_1993}. This testimony contributed to weakening legal efforts to enforce desegregation in affected school districts.

In essence, multiple scholars acting as replicating agents of the Schelling model relate its results to real-life scenarios by arguing that the model ``reproduces'' and ``helps explaining'' segregation patterns seen in contemporary cities. There is a fallacy underpining these arguments, rarely made explicit, but persistently lingering between the lines: the assumption that because the Schelling model \textit{produces} segregation-like patterns, it must logically follow that the model explains real-world segregation. This is an erroneous conflation of model output with causal inference. Schelling never suggested that this wrong association be made, and explicitly mentioned that governmental policies and socioeconomic inequalities play a much bigger role for segregation outcomes than individual preferences do \cite{schelling_models_1969, massey_american_1993}. 

\begin{figure}[h]
\centering
\includegraphics[width=0.9\textwidth]{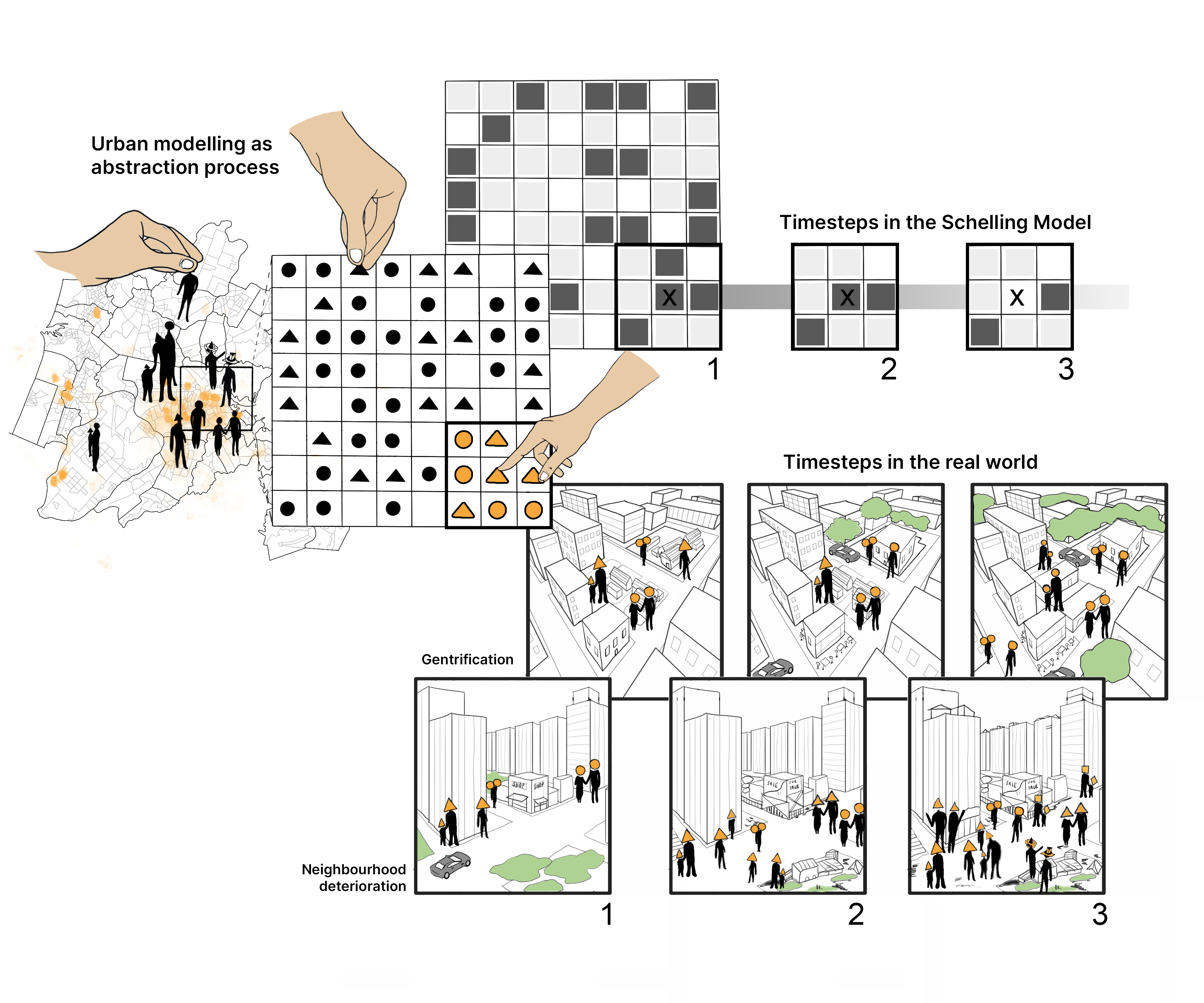}
\caption{The Schelling model as example of an urban modelling abstraction process leading us astray by obscuring the relative relevance of different process drivers. In the model, both the changes from timestep to timestep and the agents' ``satisfaction'' \cite{schelling_dynamic_1971} are locally fully determined by the ratio of ingroup vs.~outgroup agents. In the real world, processes such as \textbf{gentrification} (upper row) and \textbf{neighbourhood deterioration} (lower row), which often go hand in hand -- happening in the same city at the same time, but in different locations -- are governed by a wide range of intersecting drivers; and so are people's emotions about their neighbourhood, too.}
\label{fig:timesteps}
\end{figure}

To reframe the takeaway from the Schelling model: In a hypothetical world free of systemic oppression, legal barriers, or economic disparity, even a slight in-group preference (such as racism, classism, or religious discrimination) could lead to spatial clustering of same-group agents on a grid, and perhaps in a city. However, in our world, systemic factors far outweigh the role of individual choice (see Figure~\ref{fig:timesteps}). The continued, uncritical application of the Schelling model in segregation discourse reinforces an individualistic bias, sidelining the structural forces that shape segregation in the first place.

\subsection{Machine Learning and Forecasting}

The latest wave of quantitative segregation research has been driven by large-scale statistical modeling and Machine Learning (ML) techniques, particularly in response to the growing availability of data-driven infrastructure, evolving governance models, and policy-making initiatives \cite{delfos_perceived_2022}. ML methods have gained popularity as tools for quantifying segregation, whether through the development of multidimensional segregation indices \cite{olteanu_multidimensional_2020} or by tracking the temporal evolution of spatial demographic clustering in urban environments \cite{masias_h_evolving_2024}. Other studies have leveraged ML to explore how segregation correlates with urban form \cite{salazar_miranda_shape_2020} or the appreciation of property values in different income brackets \cite{hipp_recipes_2017}.

A particularly intriguing trend in this space is the application of ML models to longitudinal data for predictive purposes. Several recent studies have attempted to forecast the relocation patterns of future urban residents at finer spatial scales, moving beyond neighborhood-level analysis toward highly granular predictions \cite{reades_understanding_2019, stepinski_machine-learning_2022, georgati_disaggregation_2023}. This approach builds on the legacy of large-scale population prediction models \cite{robinson_machine_2018, rikani_global_2021}. A shared characteristic of these studies is their reliance on racial and ethnic attributes as key predictive variables, alongside an explicit intent of using these forecasts to inform policy-making strategies aimed at addressing segregation \cite{ko_meeting_2018}. 

One especially consequential domain relating to segregation, where fine-grained geospatial data plays a significant role, is predictive policing. Such models ingest neighbourhood characteristics and demographic data as relevant input parameters. These models use training data from real-world conditions where law enforcement patterns are deeply skewed. In contexts such as the US, certain neighborhoods and racial or ethnic groups are overpoliced and overreported as criminal threats, creating a data feedback loop that entrenches preexisting racial biases into datasets \cite{lum_predict_2016}. 

Predictive policing models fail to account for the racialized power structures embedded in their underlying data ontologies and model assumptions \cite{goodson_examining_2021}. Disregarding these systemic imbalances means that the deployment of predictive policing tools does not only reflect spatial inequalities, it actively reproduces and amplifies them \cite{jefferson_predictable_2018, jefferson_policing_2018}. This phenomenon underscores a broader ethical concern: while ML promises neutrality and precision, its application in segregation research and policy-making risks codifying social inequities into ostensibly objective, data-driven decisions.

\section{Spheres of Perpetuation of Individualistic Bias}
\label{sec:spheres}

The persistence of spatial segregation is explained through narratives that focus on personal choice, cultural differences, or individual shortcomings, while minimizing or outright ignoring the structural conditions that enable and sustain it. This tendency, theindividualistic bias, is not accidental. It is an outcome of interwoven societal processes that frame inequality as a matter of personal agency in lieu of systemic constraints.

\begin{figure}[ht]
\centering
\includegraphics[width=\textwidth]{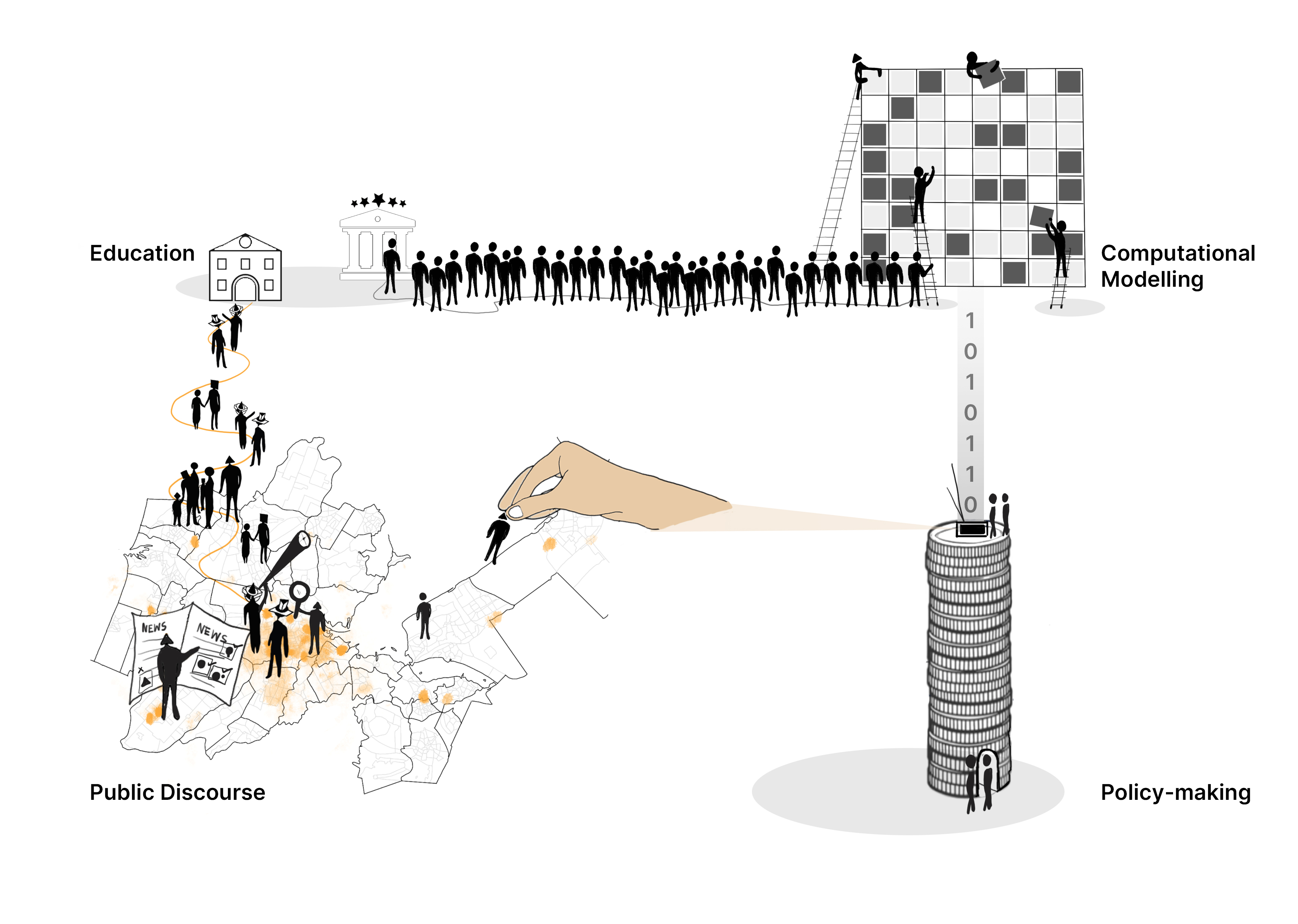}
\caption{The interconnected four spheres of perpetuation of individualistic bias: \textbf{education} often remains an elitist project, providing a supply chain of predefined mindsets to the spheres of \textbf{computational modelling} and \textbf{policy-making}; policy-making informs power brokerage that further determines peoples' livelihoods within a city; whether voices are amplified or silenced in \textbf{public discourse} is contingent on the authority these voices are granted through the other spheres.}
\label{fig:spheres}
\end{figure}

Individualistic bias does not operate in isolation. It is cultivated, reinforced, and disseminated through multiple spheres of influence, masking, legitimizing, and ultimately, perpetuating systemic inequalities. In this section, we examine four key spheres of perpetuation that uphold this bias in segregation research and discourse: computational modelling within the Engineering-Economy Complex (EEC), public discourse, education, and policy-making. These spheres interact in ways that mutually amplify each other's influence (see Figure~\ref{fig:spheres}). Knowledge produced in one sphere informs epistemological co-creation or decision-making in another, shaping both the way we understand segregation and the policies we design to address it.

Importantly, individualistic bias within these spheres is rarely acknowledged as an ideological stance. Frequently, it is embedded as tacit knowledge, subtly shaping research questions, framing public narratives, structuring curricula, and informing governance decisions. These mechanisms need to be unmasked, so we can begin to critically assess the ways in which segregation is understood, studied, and addressed, pushing beyond the simplistic focus on individual behavior toward systemic solutions that confront the structural roots of inequality.

\subsection{Computational Modelling}

Simple models spread fast. Their appeal is obvious: they are easy to replicate, easy to scale, and easy to apply across different contexts. This is also true in urban modeling, where the ability to generalize a computational framework is valued over the accuracy of its assumptions \cite{cottineau_role_2024}. When lacking a strong theoretical grounding, simple models are also among the riskiest to use in real-world decision-making. Recall the anecdote from the beginning of this essay -- the lecturer promoting the Schelling model with the assertion: ``It is simple, and it works.'' But does it? The ease with which a computational model can be removed from its original context, transferred, and applied elsewhere, makes it an attractive tool. In-depth sociological case studies, in contrast, are usually non-transferrable. Does this mean they do not ``work''? One way to determine whether a model ``works'' is to focus on what we want to do with the results. If we want to see categorical clusters -- then yes, the Schelling model ``works''. But if we want to \textit{explain} what produces comparable clusters in real-life cities, then the Schelling model \textit{does not work at all}, since it is not designed to include any of the systemic drivers of segregation.

Computational models of spatial segregation frequently borrow their workings from mainstream economics, a field that excels at simplifying human behavior into tidy, quantifiable assumptions. Simplicity plays a crucial role here, too: it is much easier to model humans as isolated data points with disconnected identities, unmoored from any sense of shared social, economic, and cultural relationships through time and space. The mathematical machinery of economics is built for this -- it assumes that humans are self-interested, fully rational, and constantly optimizing for personal gain. In turn, this promotes a vision of society where individual choices are seen as the primary drivers of outcomes \cite{hickel_less_2020, urbina_critical_2019}.

This is precisely where individualistic bias seeps in. When people are modeled as isolated decision-makers, as opposed to participants in broader systems of power, the resulting conclusions will naturally reinforce an atomistic worldview. The logic is circular: if a model is built on the premise that individual choices explain segregation, then it will inevitably find that individual choices explain segregation. The structure of the model itself precludes any serious engagement with systemic drivers like discriminatory housing policies, labor market inequalities, or the legacy of racialized urban planning from Cape Town to Rotterdam \cite{kirk_making_2019, uitermark_gentrification_2007}.

The reception history of the Schelling model is a case in point. Half a century after its conception, new versions of the model continue to emerge within computational social science. Many of these updates add technical complexity: introducing new rules, refining agent behavior, or applying machine learning techniques. But the fundamental assumption remains unchanged: segregation is portrayed primarily as a function of individual choices \cite{forse_retour_2019}. Similarly, machine learning models used to forecast segregation trends push the boundaries of computational sophistication while failing to question the ethical and methodological implications of their input variables. Race and ethnicity, often used as predictors in these models, are treated as if they were inherent, static properties of individuals, and not fluid social constructs shaped by tragedies past and present \cite{georgati_modeling_2024}. In these cases, data-driven sophistication serves to mask the real drivers of segregation.

Computational models do not just shape our understanding of segregation, they also shape which types of knowledge are taken seriously. Models from economics, engineering, and physics are granted more legitimacy than those from social sciences, simply because they offer numerical outputs that \textit{appear} objective \cite{cottineau_role_2024}. This bias creates a hierarchy of credibility, where quantitative approaches are seen as inherently superior to qualitative, context-sensitive analyses. The impact of this bias extends beyond academia. Governments and funding bodies increasingly favor data-driven approaches over qualitative studies when allocating resources for policy research \cite{solovey_social_2020, rahkovsky_ai_2021}. Within the EEC, this dynamic feeds into a self-reinforcing loop: computational models are assumed to be universally applicable, so they are applied widely. Their widespread use then serves as evidence of their utility, further entrenching their perceived authority. This modeling optimism, which is part of a broader techno-optimist mindset \cite{konigs_what_2022}, continues to spread across disciplines and through engineering and policy-making curricula.

The EEC is granted expert status in discussions on segregation, not necessarily because its models are the most accurate, but because they align with a broader cultural preference for quantitative, data-driven solutions. At the same time, the EEC draws legitimacy from public discourse and policy-making, creating a circular logic that reinforces its influence. For example, segregation research starts by asserting that ``segregation in country X is a problem'' -- an appeal to public concern, but one that rarely interrogates the deeper structural causes. After running computational analyses, these studies then conclude that their findings ''can be used for policy-making'' -- an unspecified endorsement that lends the research a sense of practical relevance, even if the specific applications remain unclear \cite{liao_uneven_2024, gambetta_mobility_2023, spierenburg_characterizing_2023}. The result is a system in which models gain credibility not necessarily by their accuracy, but by their ability to circulate within networks of academia, policy, and public discourse.

The issue is not that computational modeling has no value -- it absolutely does. The problem is that in its current prevalent form, it prioritizes simplicity and transferability over depth and critical engagement. When models of segregation ignore systemic forces, they do not just produce incomplete explanations, they actively distort our understanding of the problem, reinforcing the very inequalities they claim to study. If we want to break free from this cycle, we need to fundamentally rethink how we use computational models to study segregation. This means questioning the assumptions that underlie our frameworks, incorporating critical insights from history and sociology, and resisting the urge to default to the simplest, most transferable explanations. 

\subsection{Public Discourse}

% clickbaiting dynamics
We live in an attention economy, where the for-profit media’s primary currency is engagement, measured in clicks, views, and reactions. In this economy, nuance is a liability, and emotion is capital: fear, outrage, and scandal sell far better than complexity and careful analysis. As a result, the contemporary shift from print media to algorithm-driven social media platforms has only exacerbated this trend. News is no longer primarily curated by editors aiming for balanced reporting but by engagement-maximizing algorithms that reward sensationalism over substance \cite{nixon_business_2020, hansen_attention_2024, myllylahti_attention_2018}.

% individualistic meritocracy
Like every other institution, media is susceptible to representation bias. Controversy is amplified by giving disproportionate airtime to influential voices -- pundits, politicians, and celebrities -- who hold controversial and factually incorrect views. And these amplified voices are not representative of society at large. They are those with access to platforms, with institutional backing, with social capital. And as is often the case, those who have these privileges rarely recognize them as such. Their success is framed as a result of personal merit over structural advantage \cite{rauscher_unbequeme_2024}. When influential voices -- politicians, corporate leaders, media personalities -- tell stories of their own perseverance, they reinforce the idea that social mobility is purely a function of working hard enough \cite{allcott_social_2017, shin_twitter_2022, flintham_falling_2018}. They become living examples of the meritocratic myth, helping to obscure the reality that, for every individual success story, countless others remain trapped by barriers that no amount of personal effort can overcome \cite{orfat_redaktion_heftige_2019, rosser_trumps_2018}.

% clickbaiting & meritocracy specifically applied to urban segregation
Clickbaiting dynamics and the promotion of individualistic meritocracy play out vividly in public discourse around urban segregation. Fearmongering dominates: reports highlight the risks of no-go areas and crime-ridden neighborhoods, constructing a discourse where marginalization is presented as an inherent risk to our society. The very existence of spatially concentrated disadvantaged communities is portrayed as a problem that is implicitly linked to the failures of the individuals who live there \cite{phillips_parallel_2006, risager_territorial_2023, wittstock_bildungssystem_2024}. The focus is overwhelmingly on how ethnic concentration supposedly hinders social integration \cite{lindgren_news_2009, lipsitz_how_2011, pinkster_stickiness_2020, haynes_media_2013, cernusakova_stigma_2020}. When politicians and the media discuss disadvantaged neighborhoods, the implicit assumption is that they are disadvantaged not because of economic exclusion, racist housing policies, or mismanaged public policies, but because of the \textit{people} who live there. Residents are blamed for their own marginalization, whether through their low aspirations, refusal to integrate, or general culture of poverty \cite{murray_charles_a_losing_1994, murray_bell_1996, murray_intelligence_2009}. Meanwhile, individual success stories of those who made it out are weaponized as proof that systemic barriers do not exist \cite{mark_belief_2020, vazquez_exposure_2023, ehmke_dehumanizing_2015}. This obsession fuels anxieties around immigration and segregation while ignoring the economic structures that make migration both necessary and inevitable \cite{berry_press_2015, hindmarsh_climate_2022}. A classic example of this deeply-ingrained system is the way British media has sidelined the climate crisis in favor of an unrelenting focus on the so-called ``migrant crisis'': migrants who run the capital accumulation machinery and are forced to segregate among the deprived and left-behind neighbourhoods of cities through the country. 

% other spheres
Public discourse connects to other spheres of perpetuation in various ways. It shapes and is shaped by policy-making and the EEC, forming a self-reinforcing cycle. The media sets the agenda by identifying ``catchy anxieties'' that demand attention. Once a narrative is established, whether it is the issue of migrant neighborhoods, failing schools, or urban decay, politicians respond by formulating policies that align with public concern. These policies, in turn, further legitimize the original framing. The Danish ``ghetto laws'' offer a tremendous example. The Danish government introduced annual lists labeling certain housing estates as ``ghettos'' with high geographic precision based on socioeconomic and ethnic criteria, claiming that such areas needed state interventions to ensure integration and to counteract the development of so-called ``parallel societies'' \cite{seemann_danish_2021, blankholm_aggregating_2024}. The design of those interventions fed on the pre-existing framing of urban segregation as a cultural problem than a structural one. And once the corresponding policies were enacted, the media then adopted and normalized their language -- reinforcing the very narrative that had justified the policy in the first place \cite{andersen_denmarks_2021, risager_territorial_2023, risager_rent_2022}. This is how segregation discourse is manufactured and maintained: a feedback loop where media narratives fuel policy decisions, and policy decisions, in turn, reinforce media narratives. And all the while, the core issues -- structural racism, economic inequality, exploitative housing markets -- are conveniently left out of the conversations.

% conclusion
The stories we tell about segregation, the voices we amplify, and the narratives we prioritize, all have consequences. If we continue to frame segregation as a problem of individual choice, we will continue to pursue solutions that focus on changing individuals, and continue to avoid dismantling the systems that constrain them. To change the way we talk about segregation, we need to change who gets to shape the conversation. This means elevating voices that challenge the meritocratic myth, centering the lived experiences of marginalized communities, and demanding media accountability for the narratives it perpetuates. Most importantly, it means rejecting the idea that inequality is an unfortunate but inevitable byproduct of modern society. It is not! It is a policy choice, reinforced by a tabloid media ecosystem that thrives on oversimplification and fear. 

\subsection{Education}

As members of the EEC and academics conducting research and teaching in a university setting, we are particularly interested in the feedback loop between the EEC and education. Education is usually framed as the great equalizer: a pathway to opportunity, upward mobility, and social change. In reality, however, the further one ascends the educational ladder, the more education becomes a privilege instead of a right \cite{orfield_increasingly_2014, massey_effect_2006, massey_still_2020}. Access to educating oneself is uneven, and it is shaped by economic background, race, and social capital. Those who reach higher education, particularly at elite institutions where the production line of mindsets for the EEC is deployed in full force, find themselves in increasingly homogeneous environments, surrounded by peers who have also benefitted from privilege \cite{morley_does_2021, worthington_gilded_2021, zanten_educating_2015}. 

Discussing and understanding inequalities in such academic spaces presents a profound challenge. Classrooms in universities rarely reflect the diversity of the societies they intend to study and contribute to. Discussions about inequality take place in settings where few, if any, students have actually experienced the forms of marginalization we are analyzing \cite{messner_privilege_2011, finley_difficult_2020}. A call for mitigating these inequalities thus becomes a call to help the constructed \textit{Other}, starting from the implicit privilege of one's own position \cite{schneider_social_1993}. When students from privileged backgrounds discuss urban segregation, for example, the conversation is framed in terms of helping the marginalized, and not interrogating how their own social position benefits from structural inequalities \cite{messner_privilege_2011, finley_difficult_2020}. The focus remains on how to assist than how to dismantle, stuck on benevolence instead of accountability. As we argued in the discussion of the benefits of segregation, mainstream discourse overwhelmingly frames inequality as ``bad for the poor'', without acknowledging how it is simultaneously ``good for the rich''. The same paternalistic logic is perpetuated in the classroom. This is especially true in fields dominated by the EEC, where the social sciences are often seen as an optional afterthought: at best, an interesting perspective; at worst, an ideological distraction \cite{lennon_problem_2024, black_marxs_2022, eastman_exploring_2019, douglas_engineering_2015}. Students in these disciplines are trained to believe in the power of data, models, and optimization, but rarely challenged to consider how these tools embed specific worldviews.

Take the Schelling model, for example. For many students in computational science, their first encounter with systemic segregation is not through history or sociology, but through programming a simple agent-based model. When they code it themselves and see neat bicolor clusters emerge on the screen, it feels intuitive, logical, almost inevitable. The model appears neutral, a clean and elegant demonstration of how individual choices shape collective outcomes. But what is absent from this exercise? The reality that segregation is not simply a function of individual preferences, but a product of deliberate policy choices, racialized economic structures, and historical power imbalances. Without critical engagement, students absorb an implicit lesson: segregation happens ``naturally'', and not by institutional design. Educational institutions governed by the EEC mindset offer solid computational knowledge, but disincentivize critical interrogation. This kind of training produces professionals who are adept at building models, but who have also been systemically robbed of the opportunity to interrogate their underlying assumptions. The dominant epistemology in these fields treats inequality as a problem to be solved and not as a system to be questioned. Educational material in the field encourages students to focus on engineering and techno-optimist solutions, implicitly reinforcing the idea that segregation is a technical challenge, not a political and ethical one.

Education does not just shape the EEC, it also serves as a gatekeeper for authority across all spheres of influence. Access to elite education determines who gets to speak, whose expertise is valued, and who has the legitimacy to shape public discourse and policy \cite{jensen_bridging_2015, huang_meta-analysis_2009, richardson_racial_2021}. This creates a paradox. On the one hand, education is framed as a vehicle for social mobility. On the other, those who control decision-making spaces -- whether in academia, government, or media -- are overwhelmingly drawn from the same elite institutions \footnote{For example, Oxford University has a page celebrating world leaders who are its alumni and includes prominent names that are tied to devastating consequences of dismantling public service and austerity in the UK and beyond \cite{shafiq_oxford_2024}}. The structural barriers that limit access to higher education ensure that the voices shaping policy, public discourse, and computational modeling are not representative of broader society. This means that the narratives produced in these spaces, whether about segregation or inequality more broadly, tend to reflect the perspectives of those who are least affected by these issues.

If education is to play a role in dismantling systemic inequalities, it must move beyond the passive reproduction of dominant narratives. This requires more than just increasing access to elite institutions -- these will, by definition, remain selected few. A fundamental shift in how inequality is taught and understood is absolutely necessary \cite{van_geene_reclaiming_2025}. This means actively challenging the assumption that knowledge produced in elite spaces is inherently neutral or universal, including in the social sciences. It means questioning who is in the room, whose voices are missing, and whose lived experiences are being abstracted into agents, models, and theories. It means recognizing that education is not just about accumulating knowledge, but about cultivating the ability to critically interrogate power.

\subsection{Policy-making}

In recent decades, as global inequalities have risen, many countries have undergone a neoliberal shift in their welfare policies. Nations have made it a policy choice to move away from structural interventions and toward an emphasis on individual responsibility \cite{blokland_outside_2008, hwang_racialized_2022, risager_territorial_2023}. This shift has fundamentally shaped how issues like urban segregation are framed and addressed. A core feature of this policy landscape is the assumption that disadvantaged life outcomes result from personal shortcomings like lack of ambition, poor decision-making, or even a ``culture of poverty'' \cite{small_reconsidering_2010}. This absolves institutions of responsibility, shifting the burden onto individuals to lift themselves up through personal effort in a landscape where economic participation is already rigged.

A more subtle, yet equally insidious, manifestation of the individualistic bias is the policy approach that seeks to uplift disadvantaged individuals by relocating them to environments of higher privilege, under the assumption that proximity to wealth and success will somehow translate into material and moral betterment. A well-known example is the ``Moving to Opportunity'' (MTO) program, a large-scale social experiment on poverty deconcentration. Conducted in the 1990s in the US, the MTO program tested whether relocating low-income families from disadvantaged neighbourhoods to wealthier areas would improve their socio-economic outcomes \cite{briggs_moving_2010}. Results were mixed: while some positive effects on mental health and well-being were observed, the program showed little impact on employment, income, or educational outcomes \cite{de_souza_briggs_why_2008, gennetian_long-term_2012, sanbonmatsu_long-term_2012}. What the MTO experiment ultimately revealed is that social mobility is not simply a matter of changing an individual's surroundings. Localized social ties, access to generational wealth, and other structural barriers, all play critical roles in determining life trajectories, and these factors were largely absent from the program’s design.

The MTO program is a policy approach that focuses its efforts on individual upliftment, providing people with tools to pull themselves up by their own proverbial bootstraps. Let us be clear: these are well-intended efforts and individuals may very well benefit from such policies. Crucially, such policy programs fail to address root causes which are systemic in nature \cite{shelby_integration_2014}. MTO does nothing to address the rising cost of housing, the racial wealth gap, or the displacement effects of gentrification. It reinforces the idea that inequality is best addressed through selective interventions that help a few individuals escape, instead of rolling out policies that redistribute resources or dismantle structural barriers for everyone \cite{stal_ending_2010}.

% data-driven policymaking
In contemporary policy-making, the rise of data-driven governance has a life of its own. With advances in artificial intelligence, machine learning, and predictive analytics, governments are increasingly relying on computational models to inform policy. On the surface, this shift appears to be a move toward greater efficiency, transparency, and objectivity. Under the surface, however, it reinforces existing biases under the guise of scientific neutrality. Countless researchers in AI, algorithmic bias, and fairness have systematically described the numerous biases within AI and training datasets \cite{buolamwini_gender_2018, noble_safiya_umoja_algorithms_2018, hall_systematic_2023, arora_risk_2023, shah_comprehensive_2025}.

Consider how residential segregation is framed in policy discourse. More often than not, it is treated as a problem to be solved, where the goal is to disperse low-income (and migrant/racialized) populations in ways that supposedly promote integration. This framing is deeply influenced by computational modeling, which tends to rely on race or ethnicity as primary variables without interrogating the implications of doing so \cite{jahn_measurement_1947, williams_another_1948, schelling_models_1969, spierenburg_characterizing_2023, georgati_disaggregation_2023}. Data-driven policy-making demands measurable categories -- race, income, neighborhood demographics -- while complex social relationships, histories of systemic exclusion, and intergenerational economic disadvantages are left out because they are harder to quantify or data is not easily accessible for researchers. What results is a form of policy-making that appears highly technical and apolitical but is, in fact, deeply ideological. Because these approaches appear neutral and evidence-based, they are granted far more legitimacy than qualitative, historically grounded analyses of systemic inequality \cite{cottineau_role_2024}. Thus, policy-making does not just passively reflect existing biases, it actively shapes them, while also crucially impacting all other spheres of perpetuation of individualistic bias: 

\begin{itemize}
    \item Policy $\rightarrow$ Education: The structure of education systems is determined by policy decisions. Who gets funding, what curricula are prioritized, and which institutions are granted legitimacy. This, in turn, affects who has access to the kinds of elite educational pathways that lead to positions of power.
    \item Policy $\rightarrow$ EEC: The EEC is directly shaped by policy in the form of research funding, regulatory decisions, and the legitimization of certain types of expertise over others. The mutually reinforcing relationship between the EEC and policy-making ensures that economic and engineering approaches to social issues remain dominant.
    \item Policy $\rightarrow$ Public Discourse: Policy shapes public narratives by reinforcing the logic of individual responsibility. When policies focus on individualized solutions like relocation or selective educational interventions, they strengthen the perception that addressing inequality is a matter of personal effort and not systemic design.
\end{itemize}

The dominance of individualistic bias in policy-making is not accidental. It is deeply embedded in the neoliberal logic that has shaped governance over the past several decades. If we are serious about addressing systemic inequalities, we must confront policies that merely provide selective escape routes for a few, to truly tackle underlying structures in deliberative ways that empower us to see each other as part of a collective social frabric.

\section{You Don't Have to Live Next to Me}

In this essay, we have critically examined EEC approaches to urban segregation from within, questioning the assumptions that shape computational and policy-driven understandings of spatial inequality. We have argued that individualistic bias -- the tendency to overfocus on personal agency while diverting attention from the systemic power structures that shape life outcomes -- fundamentally hinders efforts toward meaningful structural change. Articulating segregation as the spatial manifestation of deeper social inequalities, we have laid bare an intellectual frustration: longer-term ambitions to mitigate these inequalities need systemic scrutiny and interventions. Computational approaches, however, very often tend to do the opposite.

Framing spatial inequalities in relational terms, we can better understand how the advantages accrued by some are tragically linked to the disadvantages faced by others. This perspective invites us to move beyond individualistic explanations and instead focus on the systemic structures that perpetuate inequality. Such a view not only clarifies the mechanisms of spatial injustice but also -- we hope -- fosters a sense of shared responsibility and collective action. As computational and quantitative scientists, who operate mostly in silos of EEC, we have a critical role in modeling and analyzing these vicious dynamics to carefully link root causes to actionable interventions and solutions. A shift in this perspective within our modelling processes might help us build alliances across our divided communities and work toward solutions that address the root causes of inequality. We believe that this collaborative effort is essential for creating a more equitable and just society for all, and we as computational scientists have a part to play in it. 

We have thus arrived at an urgent imperative for ourselves and for our colleagues: when choosing to develop a computational approach to an issue like urban segregation, one must engage with the social and political context in which such inequalities emerge. A critical exploration of one's own positionality is a necessary starting point and an integral element of the analysis. This could start with the following questions: What is the problem I am trying to solve? What is the evidence that it is, in fact, a problem? For whom is it a problem? Who benefits from maintaining the status quo? Which social groups have I defined as part of the problem, and why? How have I identified these groups, and how have I positioned them in opposition to one another? What assumptions am I making about people's choices, constraints, and opportunities? How do these assumptions shape the design of my model? How do I evaluate a model's outcome as positive or negative? And most critically: What kind of change do I actually want to see in spatial inequalities?

Our own engagement with these questions leads us to view urban segregation as inextricably tied to capitalist relations. In this context, individual agency, choice, or preference regarding place is not a meaningful category for analyzing structural inequalities. Computational approaches to urban segregation should aim to reduce inequalities irrespective of place. If we understand social inequalities as issues of distribution and not of representation \cite{sarbo_rassismus_2022}, then the goal should not be to prevent marginalized groups from living together \cite{shelby_integration_2014}, but to dismantle the conditions that create marginalized groups in the first place, no matter where they reside.

``You don't have to live next to me'',  Nina Simone sings in ``Mississippi Goddamn'', and continues by demanding: ``Just give me my equality''. In that simple demand lies a profound critique, transposable to the flawed assumptions underpinning contemporary policy and computational approaches to segregation. The task ahead is not to shuffle people around in the name of social mixing or integration, but to address the material conditions that entrench inequality at its root.

\section*{Artist contribution}

All images in this article have been created in a collaborative process, with visual artwork by Namrata Narendra (\url{https://namratanarendra.com}).

\section*{Acknowledgements}
For helpful comments and engaged discussions, our sincere gratitude to our dearest colleagues on this journey: Anne de Koeijer, Carolina Niewöhner, Björn Karlsson, Rainer Hegselmann, Gerwin van Schie, Martin Fleischmann, Gülşah Akçakır, Chanuwas Aswamenakul, and Dave Feldman.

\bigskip

\begin{scriptsize}
\printbibliography[heading=bibintoc]
\end{scriptsize}

% \clearpage
% \section*{Policy examples - collection}
% \input{policy-examples}

\end{document}